\documentclass[aps,english,prx]{revtex4-1}
\usepackage{amsmath}
\usepackage{SIunits}
\usepackage{graphicx}% Include figure files
\begin{document}
%
%
% Define new commands:
%
\newcommand{\ac}[0]{\ensuremath{\hat{a}_{\mathrm{c}}}}
\newcommand{\adagc}[0]{\ensuremath{\hat{a}^{\dagger}_{\mathrm{c}}}}
\newcommand{\aR}[0]{\ensuremath{\hat{a}_{\mathrm{R}}}}
\newcommand{\aT}[0]{\ensuremath{\hat{a}_{\mathrm{T}}}}
\renewcommand{\b}[0]{\ensuremath{\hat{b}}}
\newcommand{\bdag}[0]{\ensuremath{\hat{b}^{\dagger}}}
\newcommand{\betaI}[0]{\ensuremath{\beta_\mathrm{I}}}
\newcommand{\betaR}[0]{\ensuremath{\beta_\mathrm{R}}}
\newcommand{\bra}[1]{\ensuremath{\left<#1\right|}}
\renewcommand{\c}[0]{\ensuremath{\hat{c}}}
\newcommand{\cdag}[0]{\ensuremath{\hat{c}^{\dagger}}}
\newcommand{\CorrMat}[0]{\ensuremath{\boldsymbol\gamma}}
\newcommand{\Deltacs}[0]{\ensuremath{\Delta_{\mathrm{cs}}}}
\newcommand{\Deltacsmax}[0]{\ensuremath{\Delta_{\mathrm{cs}}^{\mathrm{max}}}}
\newcommand{\Deltacsparked}[0]{\ensuremath{\Delta_{\mathrm{cs}}^{\mathrm{p}}}}
\newcommand{\Deltacstarget}[0]{\ensuremath{\Delta_{\mathrm{cs}}^{\mathrm{t}}}}
\newcommand{\Deltae}[0]{\ensuremath{\Delta_{\mathrm{e}}}}
\newcommand{\Deltahfs}[0]{\ensuremath{\Delta_{\mathrm{hfs}}}}
\newcommand{\dens}[0]{\ensuremath{\hat{\rho}}}
\newcommand{\erfc}[0]{\ensuremath{\mathrm{erfc}}}
\newcommand{\Fq}[0]{\ensuremath{F_{\mathrm{q}}}}
\newcommand{\gammapar}[0]{\ensuremath{\gamma_{\parallel}}}
\newcommand{\gammaperp}[0]{\ensuremath{\gamma_{\perp}}}
\newcommand{\gavg}[0]{\ensuremath{\mathcal{G}_{\mathrm{avg}}}}
\newcommand{\gbar}[0]{\ensuremath{\bar{g}}}
\newcommand{\gens}[0]{\ensuremath{g_{\mathrm{ens}}}}
\newcommand{\gNV}[0]{\ensuremath{g_{\mathrm{NV}}}}
\renewcommand{\H}[0]{\ensuremath{\hat{H}}} 
\renewcommand{\Im}[0]{\ensuremath{\mathrm{Im}}}
\newcommand{\kappac}[0]{\ensuremath{\kappa_{\mathrm{c}}}}
\newcommand{\kappamin}[0]{\ensuremath{\kappa_{\mathrm{min}}}}
\newcommand{\kappamax}[0]{\ensuremath{\kappa_{\mathrm{max}}}}
\newcommand{\ket}[1]{\ensuremath{|#1\rangle}}
\newcommand{\mat}[1]{\ensuremath{\mathbf{#1}}}
\newcommand{\mean}[1]{\ensuremath{\langle#1\rangle}}
\newcommand{\muB}[0]{\ensuremath{\mu_{\mathrm{B}}}}
\newcommand{\omegac}[0]{\ensuremath{\omega_{\mathrm{c}}}}
\newcommand{\omegas}[0]{\ensuremath{\omega_{\mathrm{s}}}}
\newcommand{\pauli}[0]{\ensuremath{\hat{\sigma}}}
\newcommand{\pexc}[0]{\ensuremath{p_{\mathrm{exc}}}}
\newcommand{\pexceff}[0]{\ensuremath{p_{\mathrm{exc}}^{\mathrm{eff}}}}
\newcommand{\Pa}[0]{\ensuremath{\hat{P}_{\mathrm{c}}}}
\newcommand{\Qmin}[0]{\ensuremath{Q_{\mathrm{min}}}}
\newcommand{\Qmax}[0]{\ensuremath{Q_{\mathrm{max}}}}
\renewcommand{\Re}[0]{\ensuremath{\mathrm{Re}}}
\renewcommand{\S}[0]{\ensuremath{\hat{S}}}
\newcommand{\Sminuseff}[0]{\ensuremath{\hat{S}_-^{\mathrm{eff}}}}
\newcommand{\Sxeff}[0]{\ensuremath{\hat{S}_x^{\mathrm{eff}}}}
\newcommand{\Syeff}[0]{\ensuremath{\hat{S}_y^{\mathrm{eff}}}}
\newcommand{\tildeac}[0]{\ensuremath{\tilde{a}_{\mathrm{c}}}}
\newcommand{\tildepauli}[0]{\ensuremath{\tilde{\sigma}}}
\newcommand{\Tcaveff}[0]{\ensuremath{T_{\mathrm{cav}}^{\mathrm{eff}}}}
\newcommand{\Techo}[0]{\ensuremath{T_{\mathrm{echo}}}}
\newcommand{\Tmem}[0]{\ensuremath{T_{\mathrm{mem}}}}
\newcommand{\Tswap}[0]{\ensuremath{T_{\mathrm{swap}}}}
\newcommand{\Var}[0]{\ensuremath{\mathrm{Var}}}
\renewcommand{\vec}[1]{\ensuremath{\mathbf{#1}}}
\newcommand{\Xa}[0]{\ensuremath{\hat{X}_{\mathrm{c}}}}

\title{Multi-mode storage and retrieval of microwave fields in a spin ensemble}

\author{C. Grezes$^{1}$, B. Julsgaard$^{2}$, Y. Kubo$^{1,7}$, M. Stern$^{1}$, T. Umeda$^{3}$,
J. Isoya$^{4}$, H. Sumiya$^{5}$, H. Abe$^{6}$,
S. Onoda$^{6}$, T. Ohshima$^{6}$, V. Jacques$^{7}$, J. Esteve$^{8}$, D. Vion$^{1}$,
D. Esteve$^{1}$, K. M{\o}lmer$^{2}$, and P. Bertet$^{1}$}

\affiliation{$^{1}$Quantronics group, SPEC (CNRS URA 2464), IRAMIS, DSM, CEA-Saclay,
91191 Gif-sur-Yvette, France }

\affiliation{$^{2}$Department of Physics and Astronomy, Aarhus University, Ny
  Munkegade 120, DK-8000 Aarhus C, Denmark.}
	
\affiliation{$^{3}$Institute of Applied Physics, University of Tsukuba, Tsukuba 305-8573
Japan}

\affiliation{$^{4}$Research Center for Knowledge Communities, University of
Tsukuba, Tsukuba 305-8550, Japan}

\affiliation{$^{5}$Sumitomo Electric Industries Ltd., Itami 664-001, Japan}

\affiliation{$^{6}$Japan Atomic Energy Agency, Takasaki 370-1292, Japan}

\affiliation{$^{7}$Laboratoire de Physique Quantique et Mol{\'e}culaire (CNRS UMR 8537), ENS de Cachan, 94235 Cachan, France}

\affiliation{$^{8}$Laboratoire Kastler Brossel, ENS, UPMC-Paris 6, CNRS, 24 rue Lhomond, 75005 Paris, France}


\date{\today}

\begin{abstract}
A quantum memory at microwave frequencies, able to store the state of multiple superconducting qubits for long times, is a key element for quantum information processing. Electronic and nuclear spins are natural candidates for the storage medium as their coherence time can be well above one second. Benefiting from these long coherence times requires to apply the refocusing techniques used in magnetic resonance, a major challenge in the context of hybrid quantum circuits. Here we report the first implementation of such a scheme, using ensembles of nitrogen-vacancy (NV) centres in diamond coupled to a superconducting resonator, in a setup compatible with superconducting qubit technology. We implement the active reset of the NV spins into their ground state by optical pumping and their refocusing by Hahn echo sequences. This enables the storage of multiple microwave pulses at the picoWatt level and their retrieval after up to $35 \mu$s, a three orders of magnitude improvement compared to previous experiments.\vspace{1cm}
\end{abstract}
\maketitle

\section{Introduction}

The ability to store a quantum state over long times is a desirable feature in many quantum information protocols. In the optical domain, Quantum memories (QM) are necessary to implement the quantum repeaters needed for future long-distance quantum networks, and are the object of active research~\cite{Julsgaard.Nature.432.482(2004),Lvovsky.NaturePhotonics(2009),clausen.nature(2011),Damon.NewJPhys.13.093031(2011)}. Quantum memories at microwave frequencies have also become of great interest in recent years because of the development of superconducting qubits which have their resonance frequency in the GHz range, in the perspective of implementing holographic quantum computing~\cite{Tordrup.PhysRevLett.101.040501(2008),Tordrup.PhysRevA.77.020301R(2008),Wesenberg.PhysRevLett.103.070502(2009)}. For such schemes, the memory should act as an ideal multi-qubit register, able to store over long times the state of large numbers of qubits and to retrieve them on-demand.

Spin ensembles have emerged as promising candidates for a multi-mode microwave quantum memory because of their long coherence time~\cite{bar-gill_solid-state_2013,tyryshkin_electron_2012,Steger08062012} and of the multiple collective modes that a spin ensemble withstands. Existing proposals~\cite{Afzelius.NJP.1367-2630-15-6-065008(2013),Julsgaard.PhysRevLett.110.250503} (inspired by optical quantum memory protocols~\cite{Damon.NewJPhys.13.093031(2011)}) proceed in two distinct steps. First, the microwave field prepared in a well-defined quantum state $\ket{\psi}$ (for instance by a superconducting qubit) is absorbed by the spin ensemble. This generates a transverse magnetisation which decays rapidly in a time $T_2^*$ due to the spread of resonance frequencies in the ensemble. Given the weakness of the coupling constant of a single spin to the microwave field, efficient absorption requires embedding the ensemble in a high-quality factor microwave resonator in order to reach the so-called high-cooperativity regime~\cite{Kubo.PhysRevLett.105.140502(2010),Schuster.PhysRevLett.105.140501(2010),Amsuss.PhysRevLett.107.060502(2011),Ranjan.PhysRevLett.110.067004,Probst.PhysRevLett.110.157001(2013)}. The second step of the memory operation consists in retrieving the initial state, by a series of operations after which the spins emit a microwave pulse in a quantum state as close as possible to $\ket{\psi}$. In~\cite{Afzelius.NJP.1367-2630-15-6-065008(2013),Julsgaard.PhysRevLett.110.250503}, this is achieved by a Hahn-echo-like sequence consisting of two consecutive $\pi$ pulses on the spins, combined with dynamical tuning of the resonator frequency and quality factor. The maximum storage time of the memory is approximately the Hahn-echo decay time $T_2$, so that the maximal number of stored quantum states is of order $T_2 / T_2^*$, a figure which can be very large in many spin systems.

\begin{figure}[t]
  \centering
  \includegraphics[width=10cm]{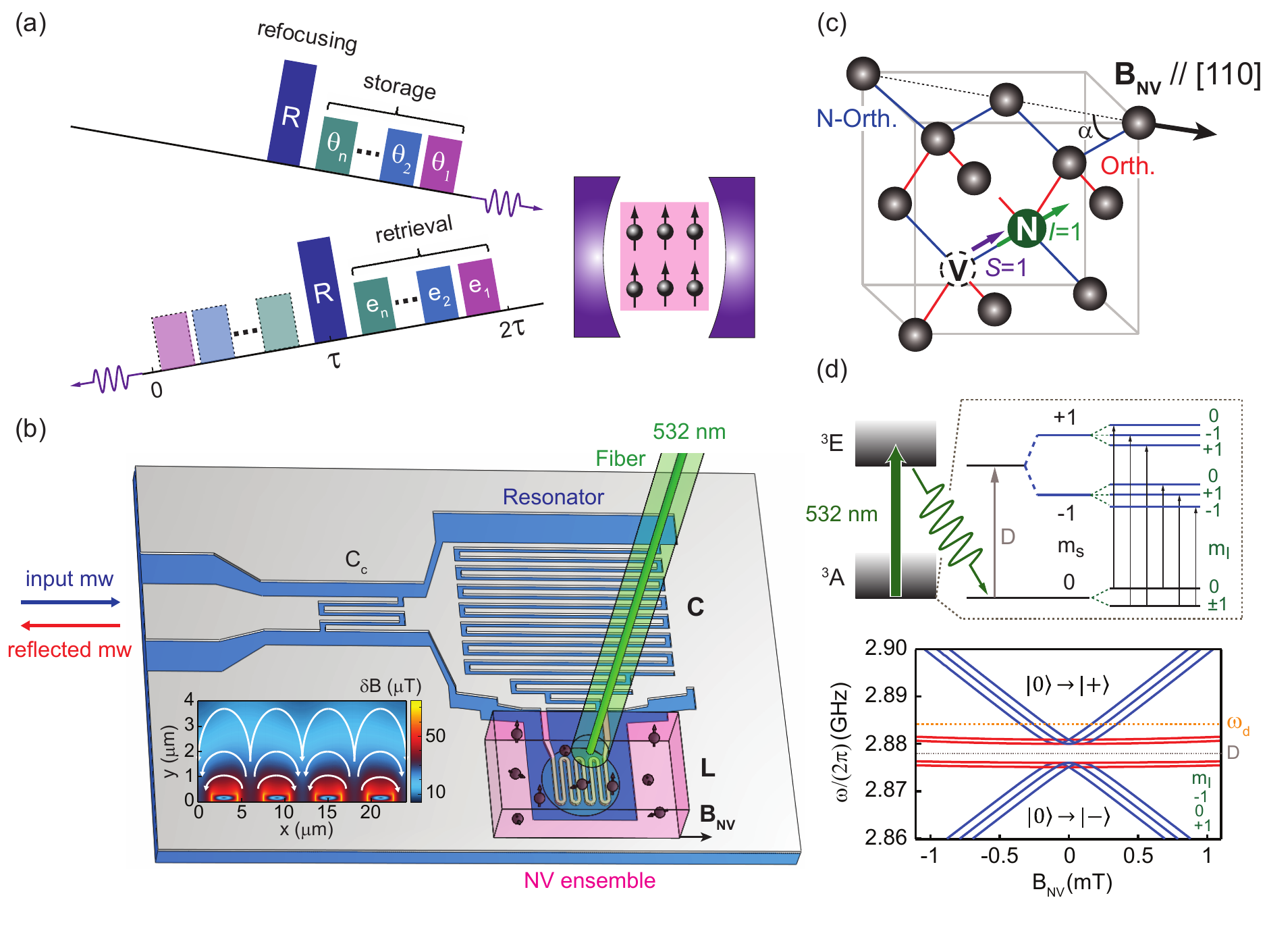}
  \caption{Principle of the experiment. (a) Scheme of the multimode two-pulse echo (2PE) protocol applied to an ensemble of spins placed in an electromagnetic cavity: successive  low-power microwave pulses $\theta_i$ are stored in the spin ensemble. A refocusing pulse $R$ acts as time-reversal for the spins and triggers the retrieval of the stored pulses as echoes $e_i$ in reverse order. Top and bottom time lines show the applied, and the reflected and echo signals, respectively. (b) Setup placed in a dilution refrigerator: the cavity is a lumped-element parallel LC resonator in niobium coupled to a coplanar waveguide by a capacitor $C_c$. It consists of an interdigitated capacitor $C$ and a meander wire inductor $L$ creating the ac magnetic field shown in inset, for a $10 \mu \mathrm{W}$ incident microwave power at resonance. The spin ensemble consists of NV centres in a diamond monocrystal pressed on top of the inductor. Laser pulses can be shone on it through an optical fibre glued to its top face. A tunable dc magnetic field $B_{NV}$ is applied parallel to the $[110]$ direction of the crystal. (c) Negatively-charged NV centres in diamond consist of a nitrogen atom next to a vacancy of the diamond lattice, having trapped an electron. Their electronic spin $S=1$ is coupled by hyperfine interaction to the nitrogen nuclear spin $I=1$ (for the $^{14}\mathrm{N}$ isotope). Half of the electronic spins (sub-ensemble denoted N-Orth in blue)  make an angle $\alpha = 35.3\,^{\circ}$ with $B_{NV}$,  whereas the other half (sub-ensemble Orth in red)  is orthogonal to the field. (d) NV simplified energy diagram (top) showing the ground $^{3}A$ and the excited $^{3}E$ electronic states as well as the Zeeman and hyperfine structure of $^{3}A$, with $D/2\pi = 2.8775$\,GHz the zero-field splitting. (bottom) Magnetic field dependence of the allowed transitions for both N-Orth (blue) and Orth (red) sub-ensembles, showing respectively a linear and quadratic Zeeman effect. NVs can be optically repumped in their $m_S=0$ ground state by application of green ($532$ nm) laser pulses exciting the $^{3}A$ - $^{3}E$ transition.}
	\label{fig1}
	\end{figure}

The first step of this protocol (quantum state transfer) has been demonstrated at the single-photon level in recent experiments~\cite{Kubo.PhysRevLett.107.220501(2011),zhu_coherent_2011}; the remaining obstacle to a microwave quantum memory is therefore the implementation of Hahn-echo refocusing sequences at the quantum level in a hybrid quantum circuit. The object of this work is precisely to identify the challenges posed by this task and to demonstrate experimentally that they can be solved. For simplicity, we consider from now on a protocol simpler than the full QM~\cite{Julsgaard.PhysRevLett.110.250503} but which constitutes an essential building block: the Two-Pulse Echo (2PE). As depicted in Fig.~\ref{fig1}a, the 2PE consists in storing weak pulses $\theta_i$ into the spin ensemble at times $t_i$, and applying a single refocusing pulse at time $\tau$ which triggers the emission of echo pulses $e_i$ at times $2 \tau - t_i$ (therefore in reverse order) in the detection waveguide~\cite{anderson:JApplPhys.26.1324(1955)}.

Performing the 2PE at the quantum level imposes a number of requirements which represent experimental challenges. For quantum states to be well defined, thermal excitations should be absent from the system. This implies both that the spin ensemble has a high degree of polarisation and that the microwave field is in its ground state with high probability, which can only be achieved if the experiments are performed at millikelvin temperatures. At these temperatures however, spins tend to relax very slowly towards their ground state, and an active spin reset is therefore needed in order to repeat the experimental sequence at a reasonable rate ($>1$\,Hz) as requested by experiments at the single photon level. Then, applying refocusing pulses to the spins requires large microwave powers potentially incompatible with the detection of quantum fields. Finally, the echo emitted by the spins should faithfully restore the initial field, which implies that the echo recovery efficiency $E$, that we define as the ratio of the energy radiated during the echo to the energy of the incoming pulse, should be close to $1$. To summarise, reaching the quantum regime requires a mean excitation per mode (both microwave and spin) $n_{mw,sp} \ll 1$, input microwave fields with intra-cavity photon number $\bar{n} \approx 1$, and an echo efficiency $E$ close to $1$.

These stringent requirements have never been met in an experiment, by far. The multi-mode character of the 2PE has been recently benchmarked in the classical regime~\cite{Wu.PhysRevLett.105.140503(2010)} with an ensemble of phosphorus donors in silicon at $10$\,K in the three-dimensional microwave cavity of an electron paramagnetic resonance spectrometer. That experiment reached $n_{mw,sp} \approx 20$, $\bar{n} \approx 10^{14}$, and an echo recovery efficiency $E \approx 10^{-10}$. Here we use negatively-charged nitrogen-vacancy (NV) centres in diamond, which are colour centres consisting of a substitutional nitrogen atom sitting next to a vacancy of the lattice (see Fig.~\ref{fig1}c) with properties suitable for a quantum memory : their spin triplet ($S=1$) electronic ground state has a long coherence time~\cite{bar-gill_solid-state_2013} and can be optically repumped in the spin ground state $\ket{m_S=0}$ (see Figs.~\ref{fig1}c and d). We re-visit the 2PE protocol with an ensemble of NV centres at $400$\,mK coupled to a planar superconducting resonator, in a setup compatible with hybrid quantum circuits, with active reset of the spin at the beginning of each experimental sequence, and we demonstrate the storage of multiple pulses at the picoWatt level for $35 \mu \mathrm{s}$, three orders of magnitude longer than in earlier experiments~\cite{Kubo.PhysRevA.85.012333}. Our experiment reaches $n_{mw} \approx 3$, $n_{sp} \approx 0.1$, $\bar{n} \approx 100$, and $E \approx 2 \cdot 10^{-4}$, and comes therefore closer to the quantum regime than previous work by several orders of magnitude. We quantitatively identify the present limitations and show that they can be solved in future experiments, opening the way to the implementation of quantum memory protocols.

\section{Experimental Setup and NV Hamiltonian}

The experimental setup is sketched in Fig.~\ref{fig1}b (see also Suppl. Info). A diamond crystal homogeneously doped with NV centres ($[NV^-]\approx 2$\,ppm) is glued on top of the inductance of a planar superconducting LC resonator cooled in a dilution refrigerator. For optical pumping, $532$\,nm laser light is injected through a single-mode optical fibre, glued on top of the crystal, $1.5$\,mm above the resonator inductance. A magnetic field $\overrightarrow{B_{NV}}$ is applied parallel to the chip along the $[110]$ crystalline axis (see Fig.~\ref{fig1}c). 

NV centres in their ground state are described~\cite{Neumann2009} by the Hamiltonian $H_{NV} / \hbar = D S_z^2 + E (S_x^2 - S_y^2) + A_z S_z I_z + \gamma_e \overrightarrow{B_{NV}} \cdot \overrightarrow{S} + Q [I_z^2 - I(I+1)/3]$, with $\overrightarrow{S}$ (resp. $\overrightarrow{I}$) the spin operator of the $S=1$ NV electronic spin (resp. the $I=1$ nitrogen nuclear spin), $D/2\pi = 2.8775$\,GHz the zero-field splitting between states $m_S=0$ and $m_S = \pm 1$, $A_z = -2.1$\,MHz the hyperfine coupling, and $Q=-5$\,MHz the nuclear quadrupole momentum~\cite{Felton.PhysRevB.79.075203(2009)}. Local electric field and strain couple with strength $E$ the spin eigenstates $\ket{m_S = \pm 1}$~\cite{dolde_electric-field_2011}. The energy eigenstates $\ket{\pm}$, shown in Fig.~\ref{fig1}d, are thus linear combinations of states $\ket{m_S = \pm 1}$; in particular, at zero magnetic field, states $\ket{\pm} = (\ket{m_S=+1} \pm \ket{m_S = -1} )/ \sqrt{2}$ are separated in energy by $2E$. In the experiment we use transitions between the spin ground state $\ket{m_S=0}$ and the two excited states $\ket{\pm}$ at frequencies close to the zero-field splitting. 

\begin{figure}[t]
  \centering
  \includegraphics[width=10cm]{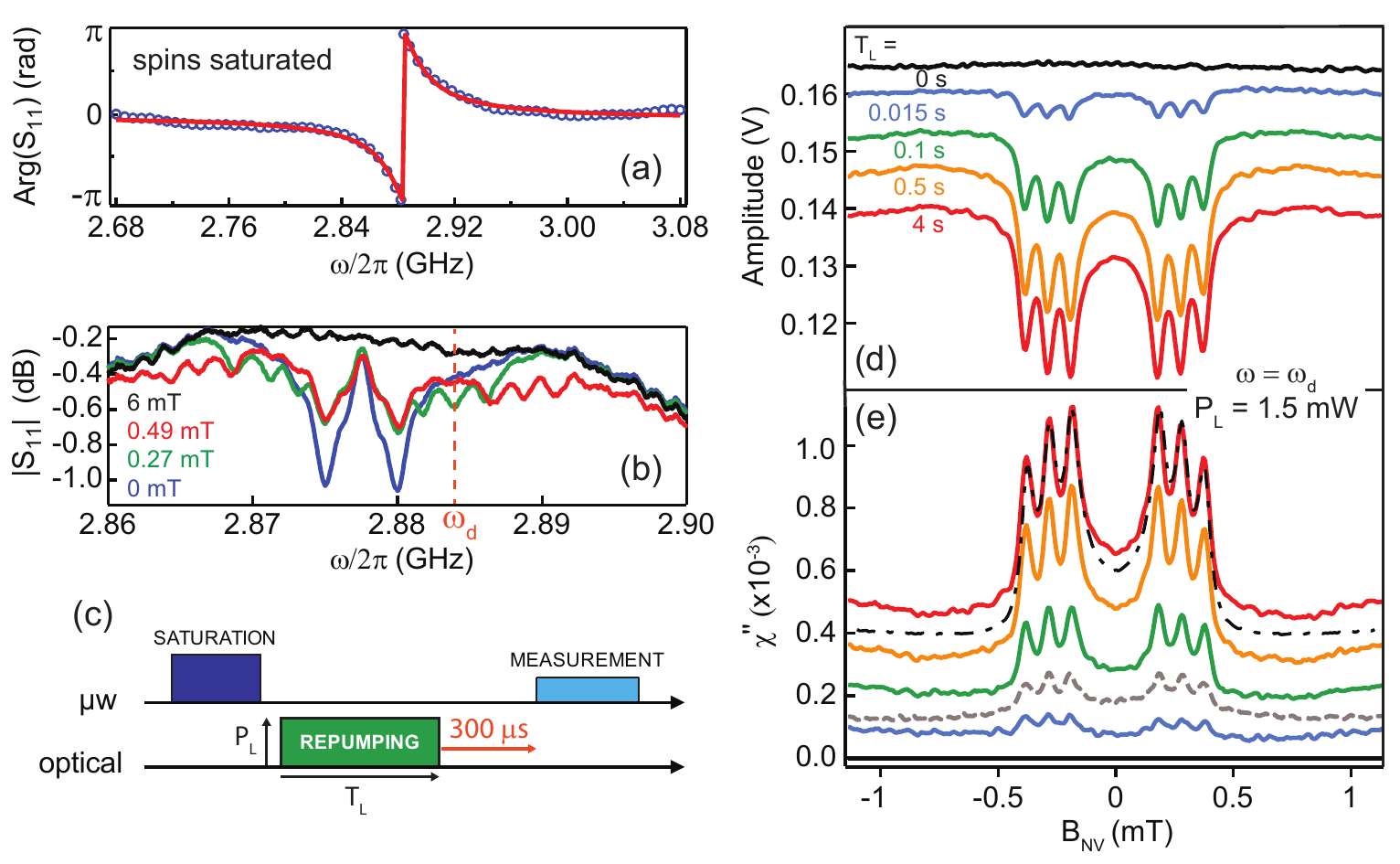}
  \caption{Spectroscopic signals and optical repumping. (a) Measured (open circles) and fitted (solid line) phase of the reflection coefficent $S_{11}$ showing the resonator resonance at $\omega_0 / 2 \pi = 2.88$\,GHz with quality factor  $Q=80$, when the spins are saturated and do not contribute to the signal. (b) Measured reflection coefficient modulus $|S_{11}|$ around the centre of the resonator line, showing the absorption by the spins for different magnetic fieds. Top line (6 mT, black) corresponds to all spins (Orth. and N-Orth) being far detuned and shows no absorption. Other lines show several absorption peaks moving with magnetic field (sub-ensemble N-Orth) or not (sub-ensemble Orth). (c) Optical reset of the NV centre spins. The spins are first saturated by a $20 \mu \mathrm{s}$ long  microwave pulse with frequency $\omega_d$ and applied power $-24$\,dBm; they are then optically repumped to their ground state with a laser pulse of power $P_L $ and duration $T_L$; after letting the system cool down during $300 \mu \mathrm{s}$, the reflected amplitude of an applied weak ($-132$\,dBm) $20$\,ms long measurement pulse at $\omega_d / 2 \pi= 2.884$\,GHz is measured. (d) Reflected amplitude for $P_L = 1.5$\,mW and different $T_L$. The curves show the hyperfine split $m_S=0$ to $m_S=\pm 1$ spectroscopic transitions of the N-Orth sub-ensemble, with an amplitude that increases with $T_L$ because of increasing spin re-polarisation. (e) Corresponding imaginary part $\chi''(B_{NV})$ of the spin susceptibility. In addition, the dashed and dash-dotted lines show respectively $\chi''(B_{NV})$ measured at thermal equilibrium ($30$\,mK, no saturating nor optical pulse) and  calculated (see Supplementary Methods and Supplementary Figs.\,$S3$ and $S4$) and rescaled by a global factor to match the experiment  at  $T_L=4$\,s.}
\label{fig2}
\end{figure}

The resonator is capacitively coupled to measurement lines through which microwave signals are applied, the amplitude and phase of the reflected field being detected by homodyne demodulation after amplification at $4$\,K. The reflection coefficient $S_{11}$, shown in Figs.~\ref{fig2}a and b, yields the resonator frequency $\omega_c/2\pi = 2.88$\,GHz and quality factor $Q=80$. Such a low $Q$ was chosen to avoid spin relaxation by superradiant spontaneous emission after excitation by the refocusing pulse~\cite{Julsgaard.PhysRevA.86.063810(2012)}. Dips in $|S_{11}|$ are due to absorption by the NVs, as evidenced by their dependence on $B_{NV}$.

\section{Active reset of the spins} 

To demonstrate optical repumping of the NVs in $\ket{m_S=0}$, we probe the spin polarisation after a laser pulse of power $P_L$ and duration $T_L$, by measuring the absorption of a microwave pulse at $\omega_d/2\pi = 2.884$\,GHz. In addition to repumping the spins, the laser generates quasiparticles in the superconductor and carriers in the silicon substrate. We thus introduce a delay of $300 \mu \mathrm{s}$ between the two pulses for these excitations to relax. In order to start from a reproducible spin polarisation, a strong microwave pulse is applied before the laser pulse, which saturates all the spins at the beginning of each sequence (see Fig.~\ref{fig2}c).

\begin{figure}[t!]
  \centering
  \includegraphics[width=8.6cm]{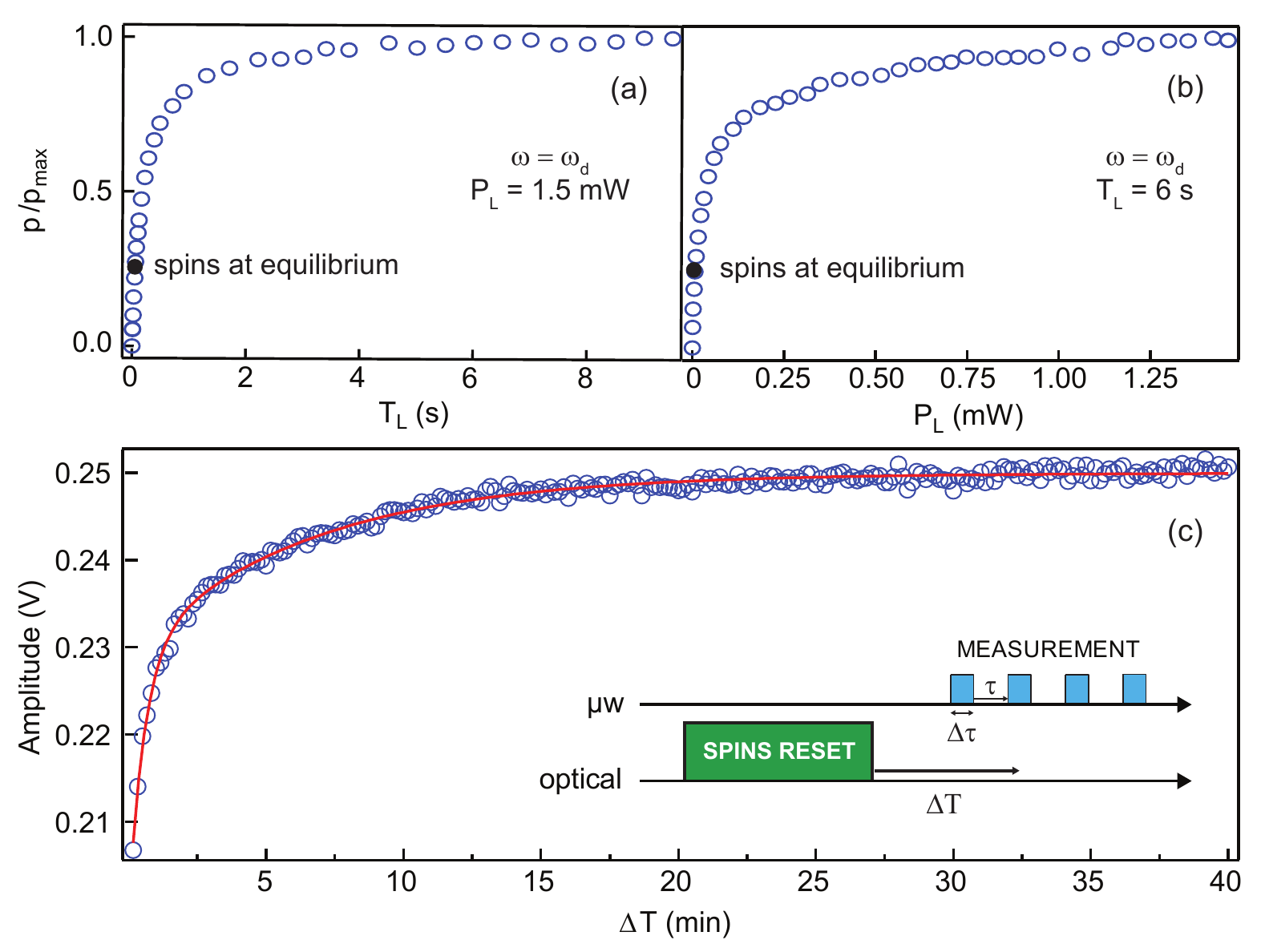}
  \caption{Spin reset efficiency and relaxation (a,b) Relative spin polarisation dependence on $T_L$ for $P_L=1.5$\,mW, and on $P_L$ for $T_L=6$\,s. The experimental sequence is shown in Fig.~\ref{fig2}c. (c) NV spin relaxation time measurement. A series of $\Delta \tau=20$\,ms weak microwave pulses ($-120$\,dBm) at $\omega_d / 2 \pi= 2.884$\,GHz, separated by $\tau=10$\,s, is applied following optical reset of the spins. Blue dots are the average reflected amplitude of each pulse. A bi-exponential fit (red solid line) yields $T_{1,a}=35$\,s and $T_{1,b}=395$\,s.}
\label{fig2b}
\end{figure}

The results are shown in Fig.~\ref{fig2}d for $P_L = 1.5$\,mW. Without laser pulse, the reflected pulse amplitude is independent of $B_{NV}$, proving that the spins are efficiently saturated by the initial microwave pulse. For non-zero $T_L$, absorption peaks with the triplet shape characteristic of the NV hyperfine structure are observed, indicating sizeable NV polarisation. To quantify the effect, we convert the absorption signal into the imaginary part of the spin susceptibility $\chi''(T_L,B_{NV})$ (see Fig.~\ref{fig2}e and Supplementary Information), which yields the relative spin polarisation $p(T_L)=\chi''(T_L,B_{NV}) / \chi''(T_{max},B_{NV})$, with $T_{max}$ the maximum repumping time. The polarisation increases with $T_L$ and then saturates (see Figs.~\ref{fig2b}a and b), which shows that the spins reach the maximum polarisation allowed by optical pumping at $532$\,nm, close to $90\%$ according to earlier work~\cite{Robledo.Nature.477.574(2011)}. The refrigerator cold stage was heated up to $400$\,mK due to laser power; all the following results were obtained under these conditions. Better alignment of the fibre with the resonator should reduce the power needed by two orders of magnitude.

Using the optical pumping, we measure the energy relaxation of the spins. In that goal the spins are first repumped, after which a series of a $20$\,ms resonant probe microwave pulse separated by $10$\,s are applied. The average reflected amplitude of each pulse is plotted in Fig.~\ref{fig2b}c and shows a bi-exponential response with time constants $T_{1,a}=35$\,s and $T_{1,b}=395$\,s, similar to recent measurements~\cite{Ranjan.PhysRevLett.110.067004}. These very long values confirm the need of actively resetting the spins for operating a QM.

\section{Pulsed response of the spins} 

\begin{figure}[t!]
  \centering
  \includegraphics[width=11cm]{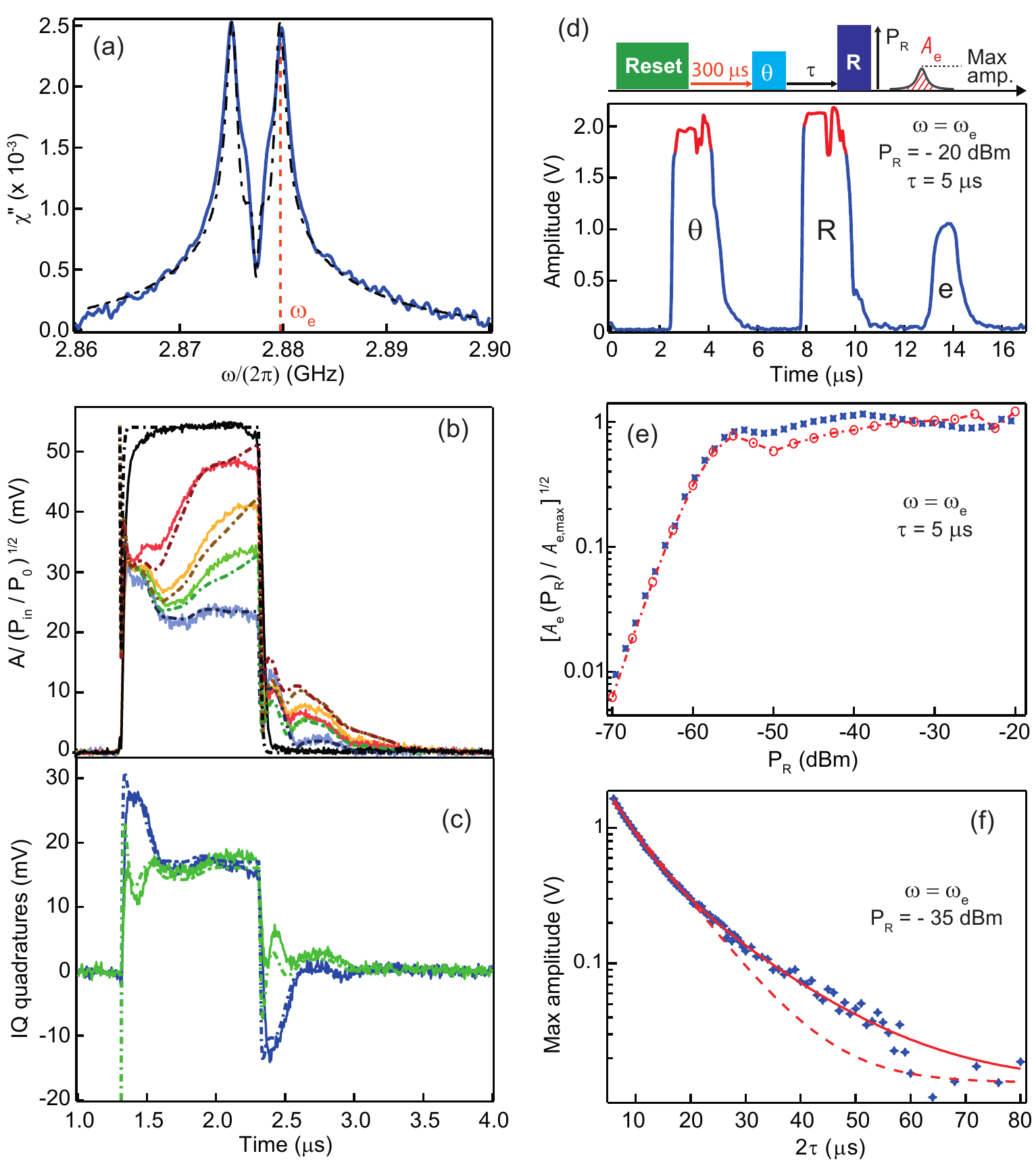}
  \caption{(a) Measured (solid line) and computed (dash-dotted line) imaginary part $\chi''(\omega)$ of the spin susceptibility at $B_{NV}=0$\,mT. The calculated curve (see Suppl. Methods) was rescaled by a global factor to match the experiment. (b) Reflected field amplitude $A$ for a square input microwave pulse of power $P_{in}$. Solid lines are experimental data with $P_{in}=-90$, $-60$, $-55$, and $-50$dBm (blue, green, yellow, and red); dashed lines are simulations. The black curve is obtained when spins have been saturated by an initial strong pulse. The curves have been rescaled by $\sqrt{P_{in}/P_0}$ for easier comparison, with $P_0 = -90$\,dBm. (c) In-phase (blue) and out-of-phase (green) quadrature of the reflected field for $P_{in}=-90$dBm. (d) Spin-echo sequence. An incoming microwave pulse $\theta$ with power $-60$\,dBm is followed by a delay $\tau$ and a $1\mu \mathrm{s}$ long refocusing pulse ($R$) with power $-20$\,dBm, yielding an echo {\it e} at time $2\tau$. Saturation of the amplifiers (shown in red) limits the measurable amplitude to about $2$\,V. (e) Experimental (crosses) and simulated and rescaled (open circles) area of the echo as a function of the refocusing pulse power $P_R$. (f) Measured (crosses) decay of the echo maximum amplitude as a function of $\tau$. Dashed and solid lines are an exponential fit yielding a characteristic time $T_{2}=8.4 \mu \mathrm{s}$  and a bi-exponential fit $f(\tau)$ yielding $T_{2A}=4.7 \mu \mathrm{s}$ and $T_{2B}=14.3 \mu \mathrm{s}$, respectively.}
\label{fig3}
\end{figure}

As a first step towards the application of refocusing pulses to the spins, we measure their time-domain response to microwave pulses of varying power. The experiments are performed at $B_{NV}=0$\,mT. The zero-field spin susceptibility $\chi''(\omega)$ (see Fig.~\ref{fig3}a) shows two broad peaks corresponding to the $\ket{0}  \rightarrow \ket{-}$ and $\ket{0} \rightarrow \ket{+}$ transitions. The width of these peaks is governed by the inhomogeneity of local electric fields and strain acting on the NVs, which results in a broad distribution of $E$, causing the hyperfine structure to be barely resolved as seen in Fig.~\ref{fig3}a. On the $\ket{0} \rightarrow \ket{+}$ transition, the spin absorption reaches a maximum at $\omega_e / 2\pi=2.8795$\,GHz, that we will thus use as the frequency of all microwave pulses in the following. Square microwave pulses of varying input power $P_{in}$ are sent to the sample, and their reflected amplitude $A$ is measured. The data are shown in Fig.~\ref{fig3}b and c, rescaled by $\sqrt{P_{in}}$, and compared to the reflected amplitude of the same microwave pulse with the spins initially saturated by a strong pulse. At low power (the linear regime), after an initial transient where resonator and spins exchange energy, $A$ reaches half of the saturated value in steady state, indicating that the spins absorb $\approx 75 \%$ of the incoming power. The steady-state value of $A$ increases with incoming power, indicating reduced spin absorption caused by progressive saturation of the ensemble. Note that no clear Rabi oscillations are observed. This is due to the spatial inhomogeneity of the microwave field generated by the planar resonator (see Fig.~\ref{fig1}b), which causes a spread of Rabi frequency within the ensemble; in particular, this prevents the application of precise $\pi$ pulses to all the spins~\cite{Malissa.RevSciInst.025116(2013)}, which is an issue for Hahn echo sequences. 

In order to understand in detail the spin dynamics, we compare the experimental data to the result of numerical simulations. These simulations consist of a number of mean value equations along the lines of~\cite{Julsgaard.PhysRevLett.110.250503} and explained in further detail in the Supplementary Information. In particular, the inhomogeneity in both spin frequency and coupling strength is taken into account by dividing the ensemble into a sufficiently large set of homogeneous sub-ensembles and integrating the equations of motion for the resonator field and the spin components of all the sub-ensembles. The distribution of spin frequencies follows from the spin susceptibility shown in Fig.~\ref{fig4}a, and the distribution of coupling strengths depend on the resonator-field vacuum fluctuations, whose spatial distribution is calculated using the COMSOL simulation package and exemplified in the inset of Fig.~\ref{fig1}b. The actual distributions used are shown in Supplementary Fig.\,S5.

The simulations employed assume an ensemble of spin-1/2 particles, which is an approximation in the case of NV centres having a spin of 1. However, in the linear, non-saturated regime this description is exact, and for the non-linear, saturated regime we expect the approximation to be justified since the applied pi pulse has a narrow frequency bandwidth and is tuned predominantly to the $\ket{0} \rightarrow \ket{+}$ transition of the NV centres. In Fig.~\ref{fig3}b and c the measured and calculated reflected field are compared and show a convincing agreement, without any adjustable parameter. This confirms the validity of the calculations, both in the linear and non-linear regime, and proves in particular that the frequency distribution used is correct.

\section{Spin-echo at high power}

Despite the impossibility to apply well-defined $\pi$ pulses to the spins, we implement a spin-echo sequence with an initial microwave pulse creating a transverse magnetisation, followed after $\tau$ by a refocusing pulse. Its power $P_R=-20$\,dBm is chosen such that spin saturation is reached within the pulse duration, as requested for spin-echo. The reflected signal amplitude is shown in Fig.~\ref{fig3}d, with the expected spin-echo observed at $2\tau$. We have studied the amplitude of this echo as a function of $P_R$, and compared this curve to the result of the simulations. The agreement is quantitative, as shown in Fig.~\ref{fig3}e; in particular the power at which the echo amplitude saturates is well predicted by the simulations. This brings further evidence of the validity of calculated coupling strengths and of the spin-1/2 approximation.
%; in particular the power at which the echo saturates is well predicted%

The dependence of the echo amplitude on $\tau$ is fitted by a bi-exponential function $f(\tau) = A \exp(- 2\tau / T_{2A}) + B \exp(- 2\tau / T_{2B})$, with two different coherence times $T_{2A} = 4.8 \mu \mathrm{s}$ and $T_{2B} = 14.3 \mu \mathrm{s}$, and $A = 0.78$ and $B=0.22$ (see Fig.~\ref{fig3}f). Such a dependence is expected for an ensemble of NV centres in zero magnetic field. Indeed, the coherence time of NV centres is limited by dipolar interactions with the surrounding spin bath, either paramagnetic impurities (P1 centres) or $^{13}\mathrm{C}$ nuclear spins. This spin bath can be approximated as generating a fluctuating magnetic field that blurs the phase of the NV centre. In zero magnetic field, an interesting situation occurs: the nuclear spin state $m_I = 0$ becomes immune to first order to magnetic fluctuations~\cite{dolde_electric-field_2011} because of the strain-induced coupling between states $m_S = \pm 1$ which gives rise to an avoided level crossing, and thus to a transition frequency independent of magnetic field to first order (see Fig.~\ref{fig1}d). This was shown in previous work to make the free-induction decay time $T_2^*$ one order of magnitude longer in zero magnetic field~\cite{dolde_electric-field_2011}, and should equally lead to a longer Hahn echo time $T_2$. This is however not true for states with $m_I = \pm 1$, which should therefore have a shorter decoherence time $T_2$ in zero magnetic field. More details will be given in future work.

\section{Multimode 2PE protocol and discussion}

\begin{figure}[t!]
  \centering
  \includegraphics[width=11cm]{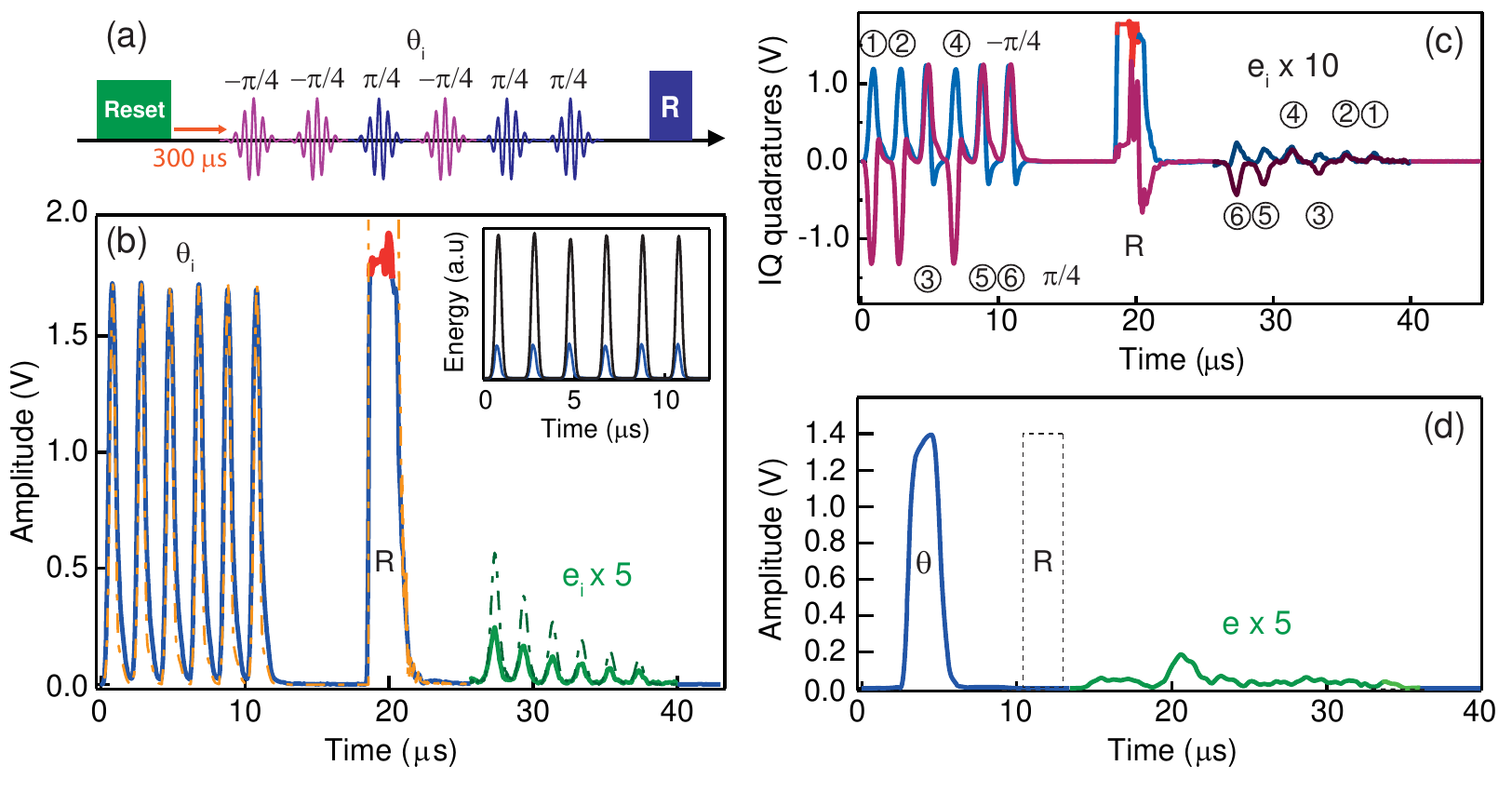}
  \caption{Test of the 2PE protocol for multimode storage of few-photon pulses. (a) Experimental sequence including a spin reset pulse, a train of six microwave pulses $\theta_i$ ($i=1,..,6$) with an identical amplitude (corresponding to $\sim 10^4$ photons in the resonator) and different phases $\varphi_1=\varphi_2=\varphi_4=-\pi/4$ and $\varphi_3=\varphi_5=\varphi_6=\pi/4$, and a $50$\,dB stronger refocusing pulse $R$ with phase  $\varphi_r = 0.1$\,rad. (b) Amplitude of the measured (solid line) and calculated (dash-dotted line) output signal showing the reflected pulses $\theta_i$ (after partial absorption by the spins) and $R$ (its amplitude being trimmed by amplifier saturation, shown in red), as well as the six re-emitted echoes $e_i$ (magnified by a factor 5). inset: The comparison between the energies of the reflected $\theta_i$ pulses with the spins saturated (black line) or reset in their ground state (blue line) shows that about 75\% of the incident power is absorbed by the spins. (c) IQ quadratures of the output signal, showing that the $e_i$ pulses (magnified by 10) are recovered with phase $-(\varphi_i - \varphi_r)$, as expected. (d) Spin-echo $e$ of $\sim 0.02$ photons in the resonator for a low power incoming $\theta$ pulse populating the resonator with only $\sim 100$ photons. The refocusing pulse (dashed line) was suppressed in the room-temperature detection chain by a microwave switch to avoid saturating the follow-up amplifiers.}
\label{fig4}
\end{figure}

We finally implement the multi-mode 2PE protocol with weak microwave pulses. Six consecutive microwave pulses with a varying phase and identical amplitude corresponding to $\approx 10^4$ photons in the resonator are first absorbed by the spin ensemble; a strong refocusing pulse is then applied $10 \mu \mathrm{s}$ later (see Fig.~\ref{fig4}a). The sequence is averaged $10^4$ times at a repetition rate of $1$\,Hz, made possible by the active reset of the spins. As shown in Fig.~\ref{fig4}b, the six pulses are recovered after the refocusing pulse up to $35 \mu \mathrm{s}$ after their storage, with an amplitude reduced by $\sim 10^2$ compared to the incoming pulse, corresponding to $\sim 1$ photon in the resonator. As expected, the pulses are re-emitted in reverse order (see Fig.~\ref{fig4}c). Note that the strong refocusing pulse ($\sim 10^9$ photons in the cavity) does not prevent detection of fields at the single-photon level few microseconds later. We were able to detect a measurable spin-echo signal for pulses containing up to $100$ times lower energy, thus populating the resonator with $\bar{n} \approx 100$ photons on average (see Fig.~\ref{fig4}d). 

An important figure of merit is the field retrieval efficiency $E$, defined as discussed in the introduction as the ratio between the energy recovered during the echo and the energy of the incoming pulse. In the data shown in Fig.~\ref{fig4}b, $E$ is seen to decrease with $\tau$ due to spin decoherence, following approximately the relation $E_e = 0.03 |f(\tau)|^2$, which yields $E=2.4 \cdot 10^{-4}$ for $2 \tau=17 \mu \mathrm{s}$. Coming back to the figures of merit defined in the introduction, our measurements reach $n_{mw} \approx 3$, $n_{sp} \approx 0.1$, $\bar{n} \approx 100$, and $E \approx 2 \cdot 10^{-4}$, many orders of magnitude closer to the quantum regime than previous state-of-the-art experiments~\cite{Wu.PhysRevLett.105.140503(2010)}. 

Reaching the quantum regime however requires a recovery efficiency $E$ close to $1$, and therefore calls for a quantitative understanding of our measurements imperfections. In that goal we have performed simulations of the multi-mode 2PE protocol. As seen in Fig.~\ref{fig4}b the measurements are well reproduced, although a seven times higher efficiency $E_t = 0.21 |f(\tau)|^2$ is predicted. We attribute the discrepancy between $E_e$ and $E_t$ to the imperfect modelling of decoherence. Indeed, our simulations treat spin decoherence in the Markov approximation. This is not an adequate treatment since it is well-known that the spin bath environment displays strong memory effects. In particular this Markov approximation is expected to describe improperly the dynamics of a spin under the action of a microwave drive, as happens during the refocusing pulse. This non-Markovian bath causes the Rabi oscillation of a single spin to decay faster than the spin-echo damping time $T_2$ as was observed in \cite{Hanson.PhysRevB.74.161203(2006)} for instance. This effect is not included in our simulations and might explain the remaining discrepancy between theory and measurements. Overall we infer from the simulations that $E_t$ would reach $0.2$ for a sample with infinite $T_2$; this number quantifies the reduced efficiency caused by refocusing pulse imperfections and finite spin absorption. In the measured efficiency $E_e \approx 2 \cdot 10^{-4}$, finite spin coherence causes a further $10^{-3}$ reduction, thus appearing as the main limitation of the field retrieval efficiency in the present experiment.

A one order of magnitude increase of the coherence time will thus be necessary to reach the quantum regime. This can be achieved~\cite{bar-gill_solid-state_2013} with samples having a reduced concentration of nitrogen paramagnetic impurities as well as isotopic enrichment of $^{12}\mathrm{C}$. Better refocusing could be obtained either by rapid adiabatic passage~\cite{Julsgaard.PhysRevLett.110.250503}, or by tailoring the spin spatial distribution~\cite{Benningshof201384(2013)}. These combined advances should make possible to reach the figures of merit requested for the quantum regime, and therefore to implement a complete quantum memory protocol~\cite{Julsgaard.PhysRevLett.110.250503,Afzelius.NJP.1367-2630-15-6-065008(2013)} at the single photon level, and to explore experimentally its fidelity. Optical pumping in a hybrid circuit, as demonstrated here, is also a first step towards the polarisation of the nitrogen nuclear spins~\cite{Jacques.PhysRevLett.102.057403}, and in a longer term towards a nuclear-spin based quantum memory.

In conclusion we have implemented the multi-mode storage and retrieval of microwave fields in an ensemble of NV centres in diamond at millikelvin temperatures, with active reset by optical pumping and refocusing by a strong microwave pulse. These results demonstrate that complex dynamical control of spin ensembles is compatible with hybrid quantum circuits, thus enabling the long-term storage of quantum information in electronic or nuclear spin ensemble quantum memory.

\textbf{Acknowledgements} We acknowledge technical support from P. S{\'e}nat, D. Duet, J.-C. Tack, P. Pari, P. Forget, as well as useful discussions within the Quantronics group and with A. Dr{\'e}au, J.-F. Roch, T. Chaneli{\`e}re and J. Morton. We acknowledge support of the French National Research Agency (ANR) with the QINVC project from CHISTERA program, of the European project SCALEQIT, and of the C'Nano IdF project QUANTROCRYO. Y. Kubo is supported by the Japanese Society for the Promotion of Science (JSPS). B. Julsgaard and  K. M{\o}lmer acknowledge support from the Villum Foundation.

\newpage

\section{Supplementary Information}

\subsection{Experimental setup and diamond sample}

The sample we use is a polished $(100)$ plate of dimensions $3\times1.5\times0.5\,\mathrm{mm}^{3}$ taken from a ${100}$ growth sector of a synthetic type-Ib diamond crystal with its edges along $[011]$. The synthetic diamond crystal was grown by a temperature gradient method under high pressure and high temperature (HPHT) of 5.5\,GPa and $1350\,^{\circ}$\,C. The crystal contained $20$\,ppm of neutral substitutional nitrogen (the P1 centre) as measured by IR absorption. Irradiation with $2$\,MeV electrons was carried out in two steps. First, it was irradiated at RT with a dose of $5\times10^{17}$\,e/c$\mathrm{m}^2$ and annealed at $800\,^{\circ}$C for $5$ hours in vacuum. Secondly, it was irradiated at $700\,^{\circ}$\,C to a dose of $5\times10^{17}$\,e/c$\mathrm{m}^2$ and annealed at $1000\,^{\circ}$\,C for $2$ hours in vacuum. From the measured absorption we deduce the NV centre concentration $\approx 2$\,ppm, implying a probable concentration of remaining neutral substitutional nitrogen (the P1 centre) of $16$\,ppm. In samples with such large P1 centre concentration, the typical NV centre coherence time is $T_2 = 5 - 10 \mu \mathrm{s}$~\cite{vanWyk.0022-3727-30-12-016(1997)}, in agreement with measurements shown in Fig.~4f.

The niobium resonator was fabricated using optical lithography followed by dry etching. Microwave simulations indicates an impedance $Z_0=\sqrt{L/C}=26 \Omega$, corresponding to a total inductance $L = Z_0 / \omega_0 = 1.4$\,nH. This inductance arises from the capacitor fingers, and from the meander wire connecting the two capacitor electrodes on top of which the diamond is pressed by a copper spring. Simulations indicate that the meander wire inductance is $L_w= 0.82$\,nH. Since the diamond crystal covers only this wire, the spin filling factor is $\eta \approx (1/2) L_w / L = 0.29$.

The wire was designed purposely to occupy a small area of $\approx 100 \times 100 \mu \mathrm{m}^2$ in order to minimise the laser power needed to repump the spins. The single-mode fibre, with numerical aperture $0.13$, was brought into our cryogen-free dilution cryostat through a home-made vacuum feedthrough. A YAG laser doubled at $532$\,nm is injected into the room-temperature end of the fibre. It is pulsed with $60$\,dB dynamics by a double-pass acousto-optic modulator. Up to $1.6$\,mW laser power could be injected into the fibre. At low temperatures, the fibre and cladding were stripped over $1$\,cm. This short bare fibre part was glued to a glass $1$\,mm thick spacer itself glued to the $0.5$\,mm thick diamond, so that the fibre - to - sample distance was $1.5$\,mm, corresponding to a nominal beam diameter of $230 \mu \mathrm{m}$ at the sample, therefore matching the area covered by the resonator meander wire. Prior to being glued, the fibre was positioned on top of this wire, with a precision estimated to be better than $0.5$\,mm.

The detailed microwave setup is shown in Supplementary Fig.~\ref{figS11}. The incoming microwave pulses are attenuated at low temperatures, routed to the input waveguide of the resonator via a circulator, and the reflected signal is amplified at $4$\,K by a low-noise HEMT amplifier, and demodulated at room-temperature, yielding the field quadratures $(I(t),Q(t))$ or equivalently the amplitude and phase $(A(t),\varphi(t))$. Note that the attenuation in the input line ($20$\,dB at $4$\,K and $10$\,dB at $100$\,mK) is not sufficient to fully suppress thermal photons in the input waveguide to the resonator, implying that a thermal field with $\approx 1$ photon is present in the resonator, and causing sizeable thermal excitation of the spin ensemble as shown in Fig.~2 of the main text. This was done purposely to apply more conveniently the refocusing pulses that require large microwave powers at the sample input. One difficulty of the experiment is to switch on and off with very high dynamics the strong microwave pulses needed to saturate or refocus the spins. We found that one microwave switch was not sufficient, and we used in all the experiments two switches in series, one internal to the microwave source, and one external (see Supplementary Figs.~\ref{figS11} and ~\ref{figS12}).

\begin{figure}[t]
  \centering
  \includegraphics[width=160mm]{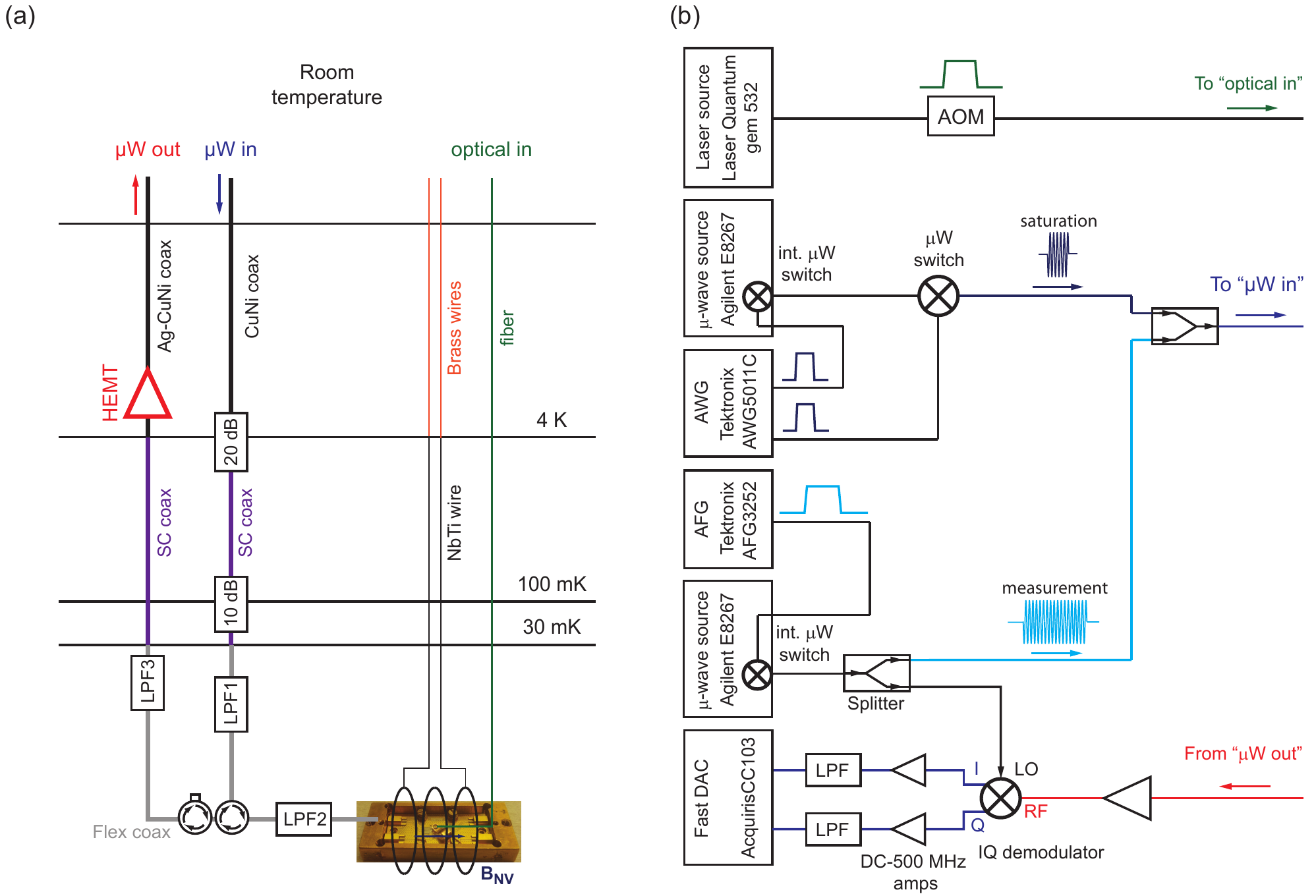}
  \caption{Measurement setup and wiring. (a) Scheme of the wiring inside the dilution refrigerator. LPF1, LPF2 and LPF3 are low-pass filters with cutoff frequencies $5.4$\,, $4.7$\, and $5.4$\, GHz, respectively. CuNi coax is a coaxial cable made of CuNi, and Ag-CuNi coax is a silver-plated CuNi coaxial cable. SC coax is a superconducting NbTi coaxial cable. Flex coax is a low-loss flexible coaxial cable. Rectangles represent ports terminated by 50$\Omega$. The cryogenic microwave amplifier is a CITCRYO 1-12 from Caltech, with gain $\sim$38 dB and noise temperature $\sim$5 K at $3$\, GHz. A DC magnetic field $B_{NV}$ is applied parallel to the chip by passing a DC current through an outer superconducting coil. The sample box and the coil are surrounded by two magnetic shieldings consisting of a lead cylinder around which permalloy tape is wrapped. The sample box, coil, and the shieldings are thermally anchored at the mixing chamber with base temperature $30$\,mK (note that in the experiments using active reset of the spins with $1.5$\,mW laser power, the temperature was $400$\,mK instead). (b) Full configuration of the measurement apparatus at room temperature for spins polarization measurements (Fig.~2 of the main text). The saturation pulse is shaped with $160$\,dB dynamics by two microwave switches in series. The DC waveform supplied to the external microwave switch is delayed by $300$\,ns to synchronize both switches. LPF is a low-pass filter with cutoff frequency $1$\, MHz. }
\label{figS11}
\end{figure}

\begin{figure}[t]
  \centering
  \includegraphics[width=160mm]{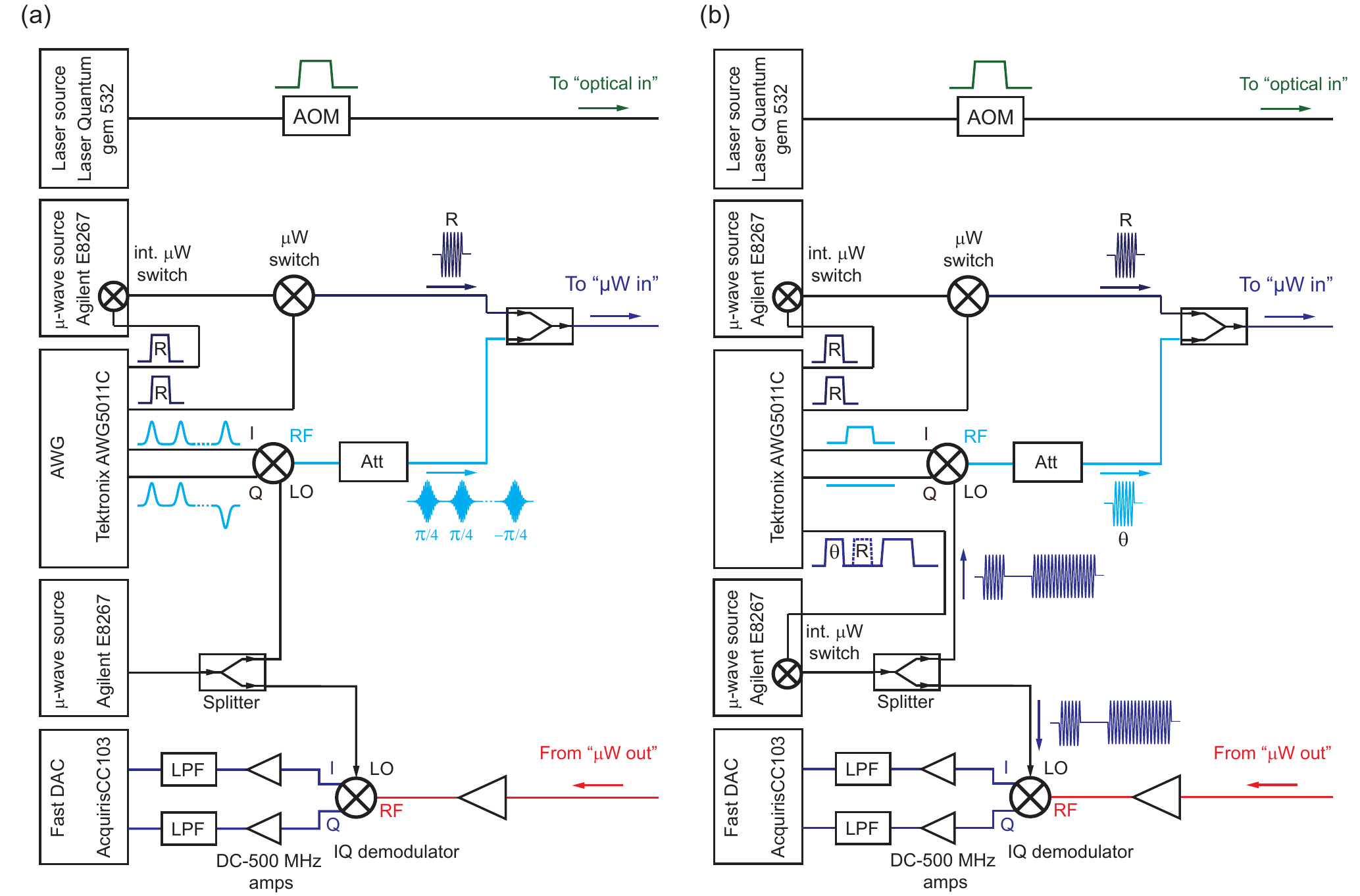}
  \caption{Detailed experimental setup for spin echo experiments. As in Supplementary Fig.\,S1, the refocusing pulse is shaped with $160$\,dB dynamics by two microwave switches in series. The DC waveform supplied to the external microwave switch is delayed by $300$\,ns to synchronize both switches. (a) Full configuration of the measurement apparatus at room temperature for multimode storage experiment (Fig.~5 b,c of main text). LPF is a low-pass filter with cutoff frequency $5$\, MHz. (b) Full configuration of the measurement apparatus at room temperature for few-photon storage experiment (Fig.~5 d of main text). The local oscillator of the IQ demodulator is pulsed to suppress the refocusing pulse from the reflected signal. LPF is a low-pass filter with cutoff frequency $1$\,MHz.}
\label{figS12}
\end{figure}

\subsection{Theory}

The goal of this section is to give the necessary elements to understand the theoretical curves presented in the main text. After defining the model, we explain 1) how the spin susceptibility $\chi''$ is extracted from microwave absorption measurements (Figs\,$2$d and $3$a in the main text), 2) how this measured susceptibility can be computed from the spin Hamiltonian assuming phenomenological distributions of the various Hamiltonian parameters (again Figs\,$2$d and $3$a in the main text), and 3) how the experimental sequences with refocusing pulses were simulated (Figs\,$4$b,c,e and $5$b of the main text).

\subsubsection{Model}

Here we follow the model already described in\,\cite{Diniz.PhysRevA.84.063810(2011),Kurucz.PhysRevA.83.053852(2011),Julsgaard.PhysRevLett.110.250503}. The spin-1 NV centers are approximated by two spin-1/2 particles (see justification in the main text and below). The NV ensemble is thus modelled as an ensemble of $N$ spin-1/2 particles of frequency $\omega_j$. Each spin couples to the cavity field (described by creation and annihilation operators $\adagc$ and $\ac$) with a coupling constant $g_j$ and a Jaynes-Cummings type of interaction. The resonator frequency is $\omega_c$, and its field damping rate $\kappa = \omega_c / 2 Q$. The total system Hamiltonian is then 

\begin{equation}
  \H = \hbar\omegac\adagc\ac + \frac{\hbar}{2}\sum_{j=1}^N\omega_j\pauli_z^{(j)}
      + i\hbar\sqrt{2\kappa}(\beta\adagc - \beta^*\ac)
    + \hbar \sum_{j=1}^N (g_j^* \pauli_+^{(j)}\ac + g_j \pauli_-^{(j)}\adagc),
\label{eq:Hamiltonian_working}
\end{equation}

with $\hat{\sigma}_k^{(j)}$ the Pauli operators of spin $j$ for $k = \{+,-,z\}$, and $\beta$ the amplitude of the microwave field driving the cavity in the laboratory frame.

The dynamics predicted by this model is quite complex (see below). However it becomes simpler in the limit where the number of excitations present in the system is much lower than the total number of spins $N$. Indeed, in this regime, spin saturation can be neglected, and the spins behave as weakly excited harmonic oscillators. This is the so-called Holstein-Primakoff approximation, by which all previous experiments on spins coupled to resonators have been theoretically described so far. The measurements shown in Figs. $2$ and $4$a of the main text are also performed in that limit, which is why we briefly discuss it in the next paragraph.

\subsubsection{Microwave absorption and spin susceptibility in the linear regime}

As shown in~\cite{Diniz.PhysRevA.84.063810(2011)} and \cite{Kurucz.PhysRevA.83.053852(2011)}, for a driving field of constant amplitude and frequency $\beta = \beta_0 e^{-i \omega t}$, the steady-state intra-cavity field amplitude is found to be $\langle a_c(t) \rangle = \langle a_c(0) e^{-i \omega t} \rangle$ with

\begin{equation}
  \langle a_c(0) \rangle = \frac{i \sqrt{2 \kappa} \beta_0}{\omega - \omega_c + i \kappa - K(\omega)},
\end{equation}
where we have introduced the function
\begin{equation}
  K(\omega) = \sum_j \frac{|g_j|^2}{\omega - \omega_j + i \gammaperp},
	\label{eq:Kdefinition}
\end{equation}

$\gammaperp=T_2^{-1}$ being the spin dephasing rate. The spins shift the resonance frequency $\omega_c$ by $Re(K)$ and add a damping term $-Im(K)$ to the field damping rate $\kappa$. These quantities can be directly extracted from the microwave measurements as explained in the following.

In the experiment, we measure the amplitude and phase of the field reflected on the resonator. We thus want to calculate the reflection coefficient $r (\omega) = \langle a_R(0) \rangle / \beta_0$, $ \langle a_R(t) \rangle = \langle a_R(0) \rangle e^{-i \omega t} $ being the reflected field. From input-output theory we have $\langle a_R(0) \rangle = \sqrt{2 \kappa}  \langle a_c(0) \rangle - \beta_0$, so that 

\begin{equation}
  r(\omega) = \frac{2 i \kappa}{\omega - \omega_c + i \kappa - K(\omega)} - 1.
\end{equation}

In the experiment, we measure reflected microwave signals through measurement cables and amplifiers which have a complex frequency-dependent transmission coefficient $T(\omega)$ giving us access to $S_{11}^*(\omega) = T(\omega) r(\omega)$ (the complex conjugate is taken because of a different sign convention between theory and experiment). To calibrate $T(\omega)$, the reflected signal $S_{11}(\omega)$ is compared to the steady-state values of the reflected signal with spins saturated which is given by $S_{11,sat}^*(\omega) = T(\omega) r_c(\omega)$, with $ r_{c}(\omega)=(\kappa+i(\omega-\omega_c))/(\kappa-i(\omega-\omega_0))$ the reflextion coefficient of the cavity without spins. In total we obtain that 

\begin{equation}
K(\omega)=\omega- \omega_c + i \kappa-i\frac{2 \kappa}{ (S_{11}^*(\omega)/S_{11,sat}^*(\omega)) r_{c}(\omega)+1}
\label{eq:Kmeasurements}
\end{equation}

We find it useful to express $K(\omega)$ in terms of the spin susceptibility $\chi(\omega)$, defined as the ratio of the induced magnetization $M_x(t)$ and the applied microwave field $H_x(t)$. More precisely for an applied field $H_x(t) = 2 H_1 \cos \omega t$, the induced magnetization is $M_x(t) = 2 H_1 (\chi'(\omega) \cos \omega t + \chi''(\omega) \sin \omega t)$, with $\chi = \chi' - i \chi''$~\cite{Abragam.NuclearMagneticResonance}. This changes the resonator inductance $L$ into $L(1 + 4 \pi \eta \chi(\omega))$~\cite{Abragam.NuclearMagneticResonance}, $\eta$ being the filling factor and $\chi$ the complex spin susceptibility in cgs units. This implies that the resonator frequency is shifted by $- 2 \pi \eta \omega_c Re(\chi)$, and the extra field damping rate is $- 2 \pi \eta \omega_c Im(\chi)$. This yields the following direct link between $K(\omega)$ and $\chi(\omega)$ : 

\begin{equation}
\chi (\omega) = - K^* (\omega) / (2 \pi \eta \omega_c). 
\label{eq:Kandchi}
\end{equation}

Equations (\ref{eq:Kmeasurements}) and (\ref{eq:Kandchi}) explain how the experimental spin susceptibility was derived from the measurements (Figs $2$d and $4$a of the main text). Note that the corresponding absorption curves were measured at powers $P \sim -132$\,dBm corresponding to few intra-cavity photons, thus by far low enough for the Holstein-Primakoff approximation to be justified.

\subsubsection{Calculation of the spin susceptibility}

The goal of this section is to demonstrate that it is possible to quantitatively understand from the NV centers Hamiltonian the measured susceptibility curves, assuming phenomenological distributions of the parameters entering this Hamiltonian. This is how we computed the theory curves in Figs.\,$2$e and $4$a of the main text. Note that this section is to a large extent independent of the rest of the paper: it explains the theory curves in Figs\,2d and inset of 3a, but importantly the numerical simulations of the echo experiments do not rely in any way on the distributions of strain or magnetic field fluctuations obtained phenomenologically in this section.

We start by rewriting the susceptibility in terms of the so-called coupling constant density function $\rho(\omega) = \sum_j |g_j|^2 \delta (\omega - \omega_j)$. From Eq.~(\ref{eq:Kdefinition}) it follows that

\begin{equation}
  K(\omega) = \int d \omega' \frac{\rho(\omega')}{\omega - \omega' + i \gammaperp}
\end{equation}

As explained in~\cite{Kurucz.PhysRevA.83.053852(2011)} this implies that $Im(K(\omega)) \approx - \pi \rho(\omega)$ (this relation holds in the limit where the inhomogeneous frequency spread is much larger than the homogeneous spin linewidth, which is the case here). Therefore, $\chi''(B_{NV},\omega)$ is proportional to $\rho(B_{NV},\omega)$. We assume that the spatial distribution (which determines the coupling constant $g_{i}$) and the frequency distribution of the spins are uncorrelated, which would be the case if the frequency distribution were only caused by local fields (magnetic, electric, strain, see below), with a spatially independent distribution. One can then write $\rho(\omega)=g_{ens}^{2}\tilde{\rho}(B_{NV},\omega)$, with $g_{ens}^2 = \sum_j |g_j|^2$ and $\tilde{\rho}(B_{NV},\omega)$ normalized such that $\int \tilde{\rho}(B_{NV},\omega) d\omega = 1$. What we are interested in here is to reproduce the frequency distribution
$\tilde{\rho}(B_{NV},\omega)$ observed in the experiment, starting from
the NV center Hamiltonian, with only one distribution of the Hamiltonian
parameters (strain $E$, magnetic field $B$, zero-field splitting
$D$).

\paragraph{NV centers distribution}

The NV center Hamiltonian (for $^{14}N$ nucleus) in the secular appoximation
is

\[
H / \hbar =DS_{Z}^{2}+E(S_{X}^{2}-S_{Y}^{2})+QI_{Z}^{2}+AI_{Z}S_{Z}+g_{NV}\mu_{B}(S_{X}B_{X}+S_{Y}B_{Y}+S_{Z}B_{Z})
\]

with $D\simeq2\pi\times2.8775$~GHz the zero-field splitting, $E$
the strain splitting, $Q=2\pi\times-5$~MHz the nuclear quadrupole
momentum, $A=2\pi\times-2.1$~MHz the hyperfine coupling of the NV
to the $^{14}N$ nucleus, and $\stackrel{\rightarrow}{B}$ the magnetic field felt by the
NV. Our ensemble of NV centers has a certain frequency distribution
because the Hamiltonian parameters have a distribution, that we assume
to be static. Here we will consider that both $A$ and $Q$ are
fixed for all NVs. On the other hand, $B_{Z}$ has evidently a certain
distribution characterized by a function $\rho_{B}(B_{Z})$ such that
the number of spins seeing a certain magnetic field between $B_{Z}$
and $B_{Z}+dB_{Z}$ is given by $N(B_{Z})=\rho_{B}(B_{Z})dB_{Z}$.
This distribution originates from the different magnetic environments
due to the local random distribution of $P1$ centers and $^{13}C$
nuclei. Note that although one can safely assume that $B_{X}$, $B_{Y}$
and $B_{Z}$ have the same distribution, we will only consider the
$B_{Z}$ distribution because it is the one that couples most strongly
to the NV center, a good approximation when $D \gg E,|g_{NV} \mu_B B|$ as is the case here. In the following we write $B \equiv B_Z$, and we note that $B=B_{NV}\cos\alpha+b$, $B_{NV}$ being the
applied magnetic field at an angle $\alpha$ from the NV axis and
$b$ the z component of the field due to the local environment of
each NV. What is constant in the problem is the distribution of $b$,
$\rho_{b}(b)$ The strain parameter $E$ has another distribution
$\rho_{E}(E)$. And finally, the zero-field
splitting $D$ is distributed with density $\rho_{D}(D)$,
which is validated by recent work~\cite{AcostaNote}.

The Hamiltonian diagonalization leads to $9$ states, corresponding
to the 3 nuclear spin states $I_{Z}=+1,0,-1$, and the $3$ NV center
states due to their spin $S=1$. This gives $6$ transition frequencies
$\omega_{m_I,\pm}[E,B,D]$. Our goal is now to express $\tilde{\rho}(\omega)$
as a function of $\rho_{b}$, $\rho_{D}$,$\rho_{E}$. We write 

\begin{equation}
\tilde{\rho}(\omega,B_{NV})=\sum_{m_I,\pm}\int\int\int dbdEdD\rho_{b}(b)\rho_{E}(E)\rho_{D}(D)\delta\left(\omega-\omega_{m_I,\pm}[E,B_{NV},D,b]\right).
\end{equation}

For $\rho_{b}$
and $\rho_{D}$ we will assume a Lorentzian shape, which at least
for $\rho_{b}$ has a physical justification (the linewidth of a dipolar
broadened spin ensemble is usually Lorentzian), with a width that
will be ``guessed'' or adapted to fit the data. For $\rho_{E}$
we use the $B_{NV}=0$ dataset (see Fig.~$4$a of the main text) to find an appropriate distribution. 

The formula above is in principle sufficient to compute $\tilde{\rho}(\omega)$
numerically given $\rho_{b}$, $\rho_{D}$,$\rho_{E}$ ; however it
would lead to very long calculation times and we need to simplify
it. The first simplification is that instead of explicitly
diagonalizing the Hamiltonian to obtain $\omega_{m_I,\pm}[E,B_{NV},D,b]$
we use approximate formulas :

$\omega_{0,\pm}[E,B_{NV},D,b]=D\pm\sqrt{E^{2}+(g_{NV}\mu_{B})^{2}(B_{NV}\cos\alpha+b)^{2}}$

$\omega_{+1,\pm}[E,B_{NV},D,b]=D\pm\sqrt{E^{2}+(g_{NV}\mu_{B})^{2}(B_{NV}\cos\alpha-B_{hfs}+b)^{2}}$

$\omega_{-1,\pm}[E,B_{NV},D,b]=D\pm\sqrt{E^{2}+(g_{NV}\mu_{B})^{2}(B_{NV}\cos\alpha+B_{hfs}+b)^{2}}$

with $B_{hfs}=\left|A/(g_{NV}\mu_{B})\right|$, considering the hyperfine interaction with the nuclear spin as a nuclear-spin-state dependent effective magnetic field of modulus  $B_{hfs}$. These formulas are valid when $D \gg E,|g_{NV} \mu_B B|$, a very good approximation in our case. This allows to very
easily invert the formula yielding, for given frequency $\omega$,
strain $E$ and zero-field splitting $D$, the local magnetic field
$b_{m_I,\pm}$ needed so that $\omega_{m_I,\pm}[E,B_{NV},D,b]=\omega$.
This equation has either zero or two solutions depending on $\omega$.
For the $0\rightarrow+$ transitions there are two solutions if $\omega\geq D+E$,
and zero else ; for the $0\rightarrow-$ transitions there are two
solutions if $\omega\leq D-E$, and zero elsewhere. 

For the $0\rightarrow+$ transitions :

$b_{0,+}^{(1)}[\omega,E,B_{NV},D]=\sqrt{(\omega-D)^{2}-E^{2}}/g_{NV}\mu_{B}-B_{NV}\cos\alpha$

$b_{0,+}^{(2)}[\omega,E,B_{NV},D]=-\sqrt{(\omega-D)^{2}-E^{2}}/g_{NV}\mu_{B}-B_{NV}\cos\alpha$

$b_{+1,+}^{(1)}[\omega,E,B_{NV},D]=\sqrt{(\omega-D)^{2}-E^{2}}/g_{NV}\mu_{B}-B_{NV}\cos\alpha+B_{hfs}$

$b_{+1,+}^{(2)}[\omega,E,B_{NV},D]=-\sqrt{(\omega-D)^{2}-E^{2}}/g_{NV}\mu_{B}-B_{NV}\cos\alpha+B_{hfs}$

$b_{-1,+}^{(1)}[\omega,E,B_{NV},D]=\sqrt{(\omega-D)^{2}-E^{2}}/g_{NV}\mu_{B}-B_{NV}\cos\alpha-B_{hfs}$

$b_{-1,+}^{(2)}[\omega,E,B_{NV},D]=-\sqrt{(\omega-D)^{2}-E^{2}}/g_{NV}\mu_{B}-B_{NV}\cos\alpha-B_{hfs}$

Identical equations apply for the $0\rightarrow-$.

Using that for any function $g(x)$ which has roots ${x_i}$ the equality $\delta(g(x)) = \sum_i \delta(x-x_i)/|g'(x_i)|$ holds, we can rewrite 

\begin{eqnarray*}
\tilde{\rho}(\omega,B_{NV}) & =& \sum_{m_I,\pm}\iiint dbdEdD\rho_{b}(b)\rho_{E}(E)\rho_{D}(D)\delta\left(\omega-\omega_{m_I,\pm}[E,B_{NV},D,b]\right) \\
& = & \sum_{m_I,\pm,i}\iint dEdD\rho_{E}(E)\rho_{D}(D)\frac{\rho_{b}\left(b_{m_I,\pm^{(i)}}[\omega,E,B_{NV},D]\right)}{\left|\frac{\partial\omega_{m_I,\pm}}{\partial b}(b_{m_I,\pm}[\omega,E,B_{NV},D])\right|}.
\end{eqnarray*}

Note that from the previous formulas it is clear that the density of NV centers at a given frequency $\omega$ can have a strong dependence on the nuclear spin state. This might explain in particular why the relative contributions of the $m_I=\pm 1$ and $m_I = 0$ to the spin echo signal at $\omega_e / 2\pi = 2.8795$\,GHz were found to be slightly different from the expected $0.66$ and $0.33$ by fitting the decoherence signal (see main text). 

A difficulty arises when $\frac{\partial\omega_{m_I,\pm}}{\partial b}$
vanishes, giving rise to a divergence. To smoothen this out, we discretize the problem
: we choose some small frequency scale $d\omega_{0}$ and we solve
the equation $\omega_{m_I,+}[E,B_{NV},D,b+db]-\omega_{m_I,+}[E,B_{NV},D,b]=d\omega_{0}$.
This equation has always two solutions, we take the Min of the two
yielding the quantity $db[E,B_{NV},D,b]$. The new formula is 

\begin{eqnarray*}
\tilde{\rho}(\omega,B_{NV})& = & \sum_{m_I,\pm,i}\iint dEdD\rho_{E}(E)\rho_{D}(D)\rho_{b} \left(b_{m_I,\pm^{(i)}}[\omega,E,B_{NV},D]\right) \\ 
&& \times db[E,B_{NV},D,b_{m_I,\pm}[\omega,E,B_{NV},D]]/d \omega_0.
\end{eqnarray*}

\paragraph{Comparison with the data}

\begin{figure}[t]
  \centering
  \includegraphics[width=120mm]{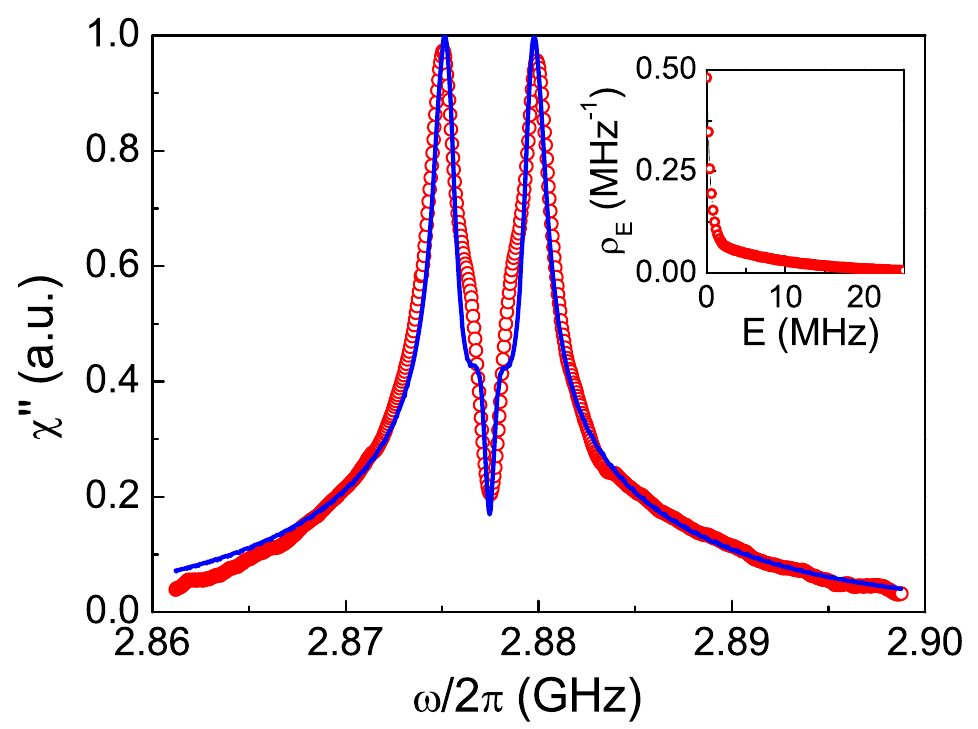}
  \caption{Rescaled spin susceptibility $\chi''(\omega,B_{NV}=0)$. Open red circles are experimental data, solid blue line is the theory computed with the bi-exponential strain distribution $\rho_{E} (E)$ shown in the inset, with a Lorentzian $\rho_b(b)$ and $\rho_D(D)$ with respective widths $dB_0=0.21$\,Gs and $dD_0/2 \pi=150$\,kHz.}
\label{figS3}
\end{figure}

We assume a Lorentzian distribution for both $\rho_{b}(b)$ and $\rho_{D}(D)$
with respective widths $db_{0}$ and $dD_{0}$. We use the
data at $B=0$ to guess the distribution $\rho_{E}(E)$.
We find that a bi-exponential distribution $\rho_{E}(E) = [\exp(-E/E_1) + A_1 \exp(-E/E_2)] / (E_1 + A_1 E_2)$ yields a computed $\tilde{\rho}(\omega,B_{NV}=0)$ that
reproduces semi-quantitatively the data. In total we use the following parameters : $db0=0.21$\,Gs, $dD_0 /2 \pi = 0.15$\,MHz, $E_1/2\pi=0.5$\,MHz, $E_2/2 \pi = 10$\,MHz, $A_1=0.2$. In this way we obtain the $B_{NV}=0$\,Gs spin susceptibility shown in Fig.~\ref{figS3} (the corresponding $\rho_{E}(E)$ distribution is shown in inset). 

\begin{figure}[t]
  \centering
  \includegraphics[width=120mm]{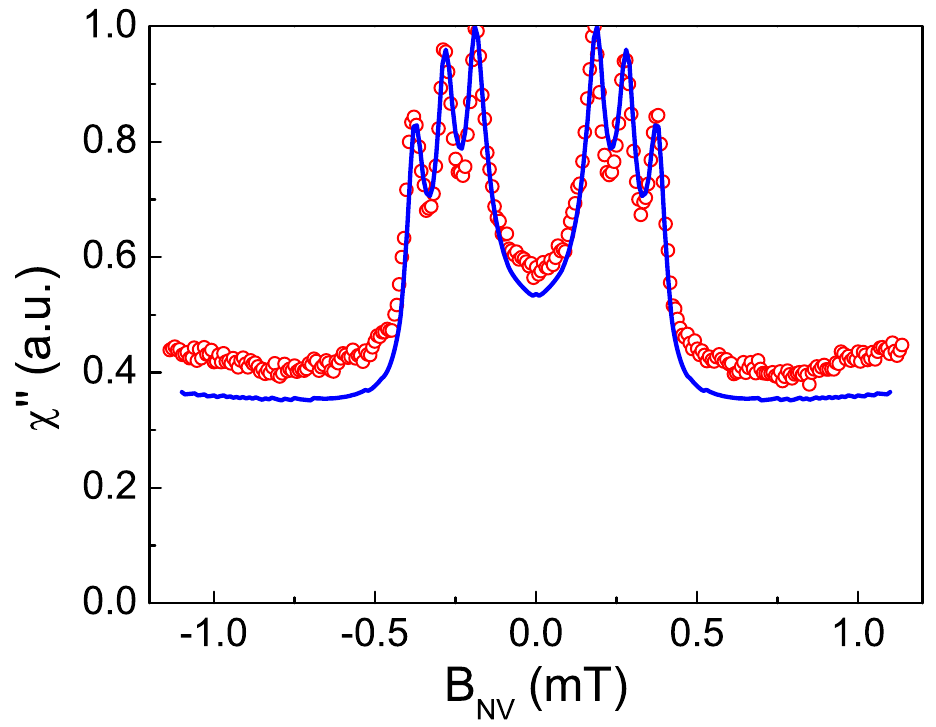}
  \caption{Rescaled spin susceptibility $\chi''(\omega_d,B_{NV})$. Open red circles are experimental data, solid blue line is the theory.}
\label{figS4}
\end{figure}

After having in this way determined the distributions $\rho_{E}(E),\rho_{D}(D),\rho_{b}(b)$, we compute without further adjustable parameters the rescaled $\chi''(\omega_d,B_{NV})$. The experimental distribution includes contributions both from the spins that are orthogonal to $B_{NV}$ and from those that are non-orthogonal, each of those having a very different resonance frequency dependence on $B_{NV}$ as shown in Fig.~2 of the main text. Each family contains exactly half of the total number of spins contributing to the signal ; however, spins from each family have a different coupling constant to the resonator field due to the angle they make with this field. This difference in coupling constant can be incorporated in a single numerical factor that yields a different ensemble coupling constant for each of the two spin families, $g_{ens,o}$ and $g_{ens,No}$. Indeed, the coupling constant of a NV center ensemble to a resonator is given by $g_{\mathrm{ens}}=g_{\mathrm{NV}}\mu_{B}\sqrt{\eta\beta\mu_{\mathrm{0}}\hbar\omega_{\mathrm{r}}(\Phi)\rho}/2\hbar$~\cite{Kubo.PhysRevLett.105.140502(2010)}, with $\eta=\int_{sample}\delta B_0^2 / \int\delta B_0^2$ the ensemble filling factor and $\beta=\int\left|\mathbf{\delta B_{0}}(\mathbf{r})\right|^{2}\sin^{2}\varphi(\mathbf{r})d\mathbf{r}/\int\left|\mathbf{\delta B_{0}}(\mathbf{r})\right|^{2}d\mathbf{r}$ the angular factor. The geometrical filling factor is clearly identical for the two families, but the factor $\beta$ differs. We have numerically calculated the ratio $\beta_{No} / \beta_{o} = 0.6$, yielding $g_{ens,No}^2 = 0.6 g_{ens,o}^2$. In this way we are able to compute the total $\chi''(\omega_d,B_{NV}) = 0.6 \chi''_{No}(\omega_d,B_{NV}) + \chi''_{o}(\omega_d,B_{NV})$. The rescaled result is shown in Fig.~\ref{figS4}.

As can be seen from Figs.~\ref{figS3} and \ref{figS4}, the agreement is semi-quantitative. All the features are reproduced but not exactly with the appropriate weight. These remaining discrepancies might be due to the fact that the strain distribution is likely to have some spatial dependence ; in particular it could depend to some extent of the distance to the surface, in which case it would be correlated with the coupling constant distribution making the above analysis an approximation. 

\subsubsection{Simulations}

Numerical simulations are performed to describe spin-echo experiments and quantitatively estimate the echo field retrieval efficiency. 
Therefore they clearly do not use the linear approximation developed in the previous paragraphs, in order to account for refocusing effects, but instead use the complete Hamiltonian Eq.~(\ref{eq:Hamiltonian_working}). To perform the calculations while taking into account the
inhomogeneity in both spin resonance frequencies and coupling
strengths, the entire inhomogeneous ensemble is divided
into $M$ homogeneous sub-ensembles, $\mathcal{M}_1, \mathcal{M}_2,
\ldots, \mathcal{M}_M$, each of them describing spins having an identical frequency $\omega_m$ and coupling to the cavity field $g_m$. The 
total number of spins in one sub-ensemble is defined as $N_m$.
We define the sub-ensemble spin collective operators:
\begin{equation}
  \label{eq:def_sub-ensembles}
  \S_x^{(m)}=\negthickspace\negthickspace\sum_{j\in \mathcal{M}_m}\pauli_x^{(j)},\quad
  \S_y^{(m)}=\negthickspace\negthickspace\sum_{j\in \mathcal{M}_m}\pauli_y^{(j)},\quad
  \S_z^{(m)}=\negthickspace\negthickspace\sum_{j\in \mathcal{M}_m}\pauli_z^{(j)}.
\end{equation}
where $m$ runs over all spin sub-ensembles. Here again, the spin-1 NV center is treated approximately as two
spin-$\frac{1}{2}$ particles. By incorporating the effect of resonator leakage and spin decoherence in
the Markov approximation, the dynamical evolution of mean values in the frame rotating at $\omega_s$ (with $\omega_s$ the mean spin frequency) is
described by (see Ref.~\cite{Julsgaard.PhysRevLett.110.250503}
for details):
\begin{align}
\label{eq:Mean_val_eq_Xa}
  \frac{\partial X_{\mathrm{c}}}{\partial t} &= -\kappa X_{\mathrm{c}}
    +\Deltacs P_{\mathrm{c}} -\sum_m \frac{g_m}{\sqrt{2}}S_y^{(m)}  
    + 2\sqrt{\kappa}\betaR, \\
\label{eq:Mean_val_eq_Pa}
  \frac{\partial P_{\mathrm{c}}}{\partial t} &= -\kappa P_{\mathrm{c}}
    -\Deltacs X_{\mathrm{c}} -\sum_m\frac{g_m}{\sqrt{2}}S_x^{(m)}  
    +2\sqrt{\kappa}\betaI, \\
\label{eq:Mean_val_eq_Sx_kl}
  \frac{\partial S_x^{(m)}}{\partial t} &= -\gammaperp S_x^{(m)}
    -\Delta_m S_y^{(m)} -\sqrt{2}g_mS_z^{(m)} P_{\mathrm{c}}, \\
   \label{eq:Mean_val_eq_Sy_kl}
  \frac{\partial S_y^{(m)}}{\partial t} &= -\gammaperp S_y^{(m)}
    +\Delta_m S_x^{(m)} - \sqrt{2}g_m S_z^{(m)} X_{\mathrm{c}}, \\
   \label{eq:Mean_val_eq_Sz_kl}
  \frac{\partial S_z^{(m)}}{\partial t} &= \sqrt{2}g_m(S_x^{(m)}P_{\mathrm{c}} 
    + S_y^{(m)} X_{\mathrm{c}}) - \gammapar(S_z^{(m)} + N_m).
\end{align}
Here $\Deltacs = \omegac - \omegas$, $\Delta_j = \omega_j - \omega_s$, $\Delta_m = \omega_m
- \omega_s$; and $\Xa = \frac{\ac+\adagc}{\sqrt{2}}$
and $\Pa = \frac{-i(\ac-\adagc)}{\sqrt{2}}$ are the cavity field quadratures such that $[\Xa,\Pa] =
i$, $\betaR$ and $\betaI$ are real and imaginary parts of
the external driving field with $|\beta|^2 = \betaR^2 + \betaI^2$
being the incident number of microwave photons per second, $\gammapar = 1/T_1$ is the
spin population decay rate. In the experiment the population decay time $T_1 \approx
35$\,s (see Fig.$3$c of main text) is very long compared to the typical refocusing time scales and
we use the excellent approximation $\gammapar = 0$.

The first step is to determine the size $N_m$ of each sub-ensemble $\mathcal{M}_m$, which requires knowledge of
the distribution of coupling constants and resonance frequency within the spin-ensemble. The distribution of coupling 
constants can be computed from the known resonator geometry and crystalline orientation (see next section); the distribution 
of resonance frequencies is not a priori known but is extracted from absorption measurements as explained earlier.

\paragraph{Determining the coupling strength distribution}
\label{sec:determ-coupl-strengt}
We first come back to the interaction Hamiltonian between the NV center spin $\vec{S}$ and the quantized
resonator magnetic field: $H_{\mathrm{I}}/\hbar =
\gNV\muB\vec{S} \cdot \delta\vec{B}(\ac +
\adagc)$, where $\delta\vec{B}$ is the rms fluctuations of the
resonator vacuum field. The external field $\vec{B}_{\mathrm{NV}}$ and
the effective field generated by the nuclear spin, $AI_Z S_Z =
\gNV\muB \frac{A
  I_Z}{\gNV\muB}S_Z \rightarrow
\gNV\muB B_{\mathrm{hfs}}S_Z$, can be treated
classically. In the rotating wave approximation the spin part of the
Hamiltonian can be expressed in terms of the energy eigen states
$\{\ket{+}, \ket{-}, \ket{0}\}$ at zero bias magnetic field as:
\begin{equation}
  \begin{split}
  H &= (D+E)\ket{+}\!\bra{+} \:+\: (D-E)\ket{-}\!\bra{-} \\
   &\quad + \: g_x \left[\ac\ket{+}\!\bra{0} \:+\: 
  \adagc\ket{0}\!\bra{+}\right] \:-\: ig_y\left[\ac \ket{-}\!\bra{0} \:-\:
   \adagc\ket{0}\!\bra{-}\right],
  \end{split}
\end{equation}
where $g_{x,y} = \gNV\muB\delta B_{x,y}$. We
note the energy splitting of $2E$ between the $\ket{+}$ and $\ket{-}$
states, and we observe that two Jaynes-Cummings-like interaction terms
emerge. Under the Holstein-Primakoff approximation (i.e.~in
absence of saturation with all population essentially in the ground
state $\ket{0}$), the NV center can be treated accurately as two
separate spin-$\frac{1}{2}$ particles. Of course, if the resonator
field saturates the NV center, and hence depletes the $\ket{0}$ state,
this two-particle picture fails. Nonetheless, we argue below that it
is a reasonable approximation to maintain this picture throughout our
simulations.

The Hamiltonian above is expressed with the quantization axis (the
$z$-axis) along the NV center axis. Since this does not coincide with
our laboratory coordinate system (defined in Fig.~1b of the main
text), we must define a local coordinate system for each of the four
possible NV center axes. As local $z$-axes we choose:
\begin{equation}
  \hat{\vec{k}}_1 =
  \begin{bmatrix}
    \sqrt{2/3} \\ \sqrt{1/3} \\ 0
  \end{bmatrix}, \quad
  \hat{\vec{k}}_2 =
  \begin{bmatrix}
    -\sqrt{2/3} \\ \sqrt{1/3} \\ 0
  \end{bmatrix}, \quad
  \hat{\vec{k}}_3 =
  \begin{bmatrix}
   0 \\ -\sqrt{1/3} \\ \sqrt{2/3}
  \end{bmatrix}, \quad
  \hat{\vec{k}}_4 =
  \begin{bmatrix}
    0 \\ -\sqrt{1/3} \\ -\sqrt{2/3}
  \end{bmatrix},
\end{equation}
i.e.~for family 1 we take, $\hat{\vec{z}}_{\mathrm{loc,1}} =
\hat{\vec{k}}_1$, etc. For symmetry reasons the coupling strengths
from family 1 and 2 turn out identical, and likewise for families 3
and 4. We thus proceed with families 1 and 3 only. As local $x$- and
$y$-axes for family 1 we have a selection of choices:
\begin{equation}
    \hat{\vec{x}}_{\mathrm{loc,1}} =
   \begin{bmatrix}
     -\sqrt{1/3}\cos\psi \\ \sqrt{2/3}\cos\psi \\ \sin\psi
   \end{bmatrix}, \qquad
  \hat{\vec{y}}_{\mathrm{loc,1}} =
   \begin{bmatrix}
     \sqrt{1/3}\sin\psi \\ -\sqrt{2/3}\sin\psi \\ \cos\psi
   \end{bmatrix}.
\end{equation}
$\psi$ being the angle between the NV axis and the direction of non-axial strain in the diamond matrix.
For all angles $\psi$ we obtain a local orthogonal coordinate system
$[\hat{\vec{x}}_{\mathrm{loc,1}}, \hat{\vec{y}}_{\mathrm{loc,1}},
\hat{\vec{z}}_{\mathrm{loc,1}}]$ with the $z$-axis pointing along the
NV center axis as required. The angle $\psi$ must be chosen such
that the Hamiltonian term $E(\S_x^2 - \S_y^2)$ describes correctly the
actual physical strain experienced by the NV center from the
surrounding host material. However, since this strain has no preferred
direction we shall later average our coupling-strength distribution over this
angle. Similarly, for family 3 we adopt the local coordinate system:
$\hat{\vec{z}}_{\mathrm{loc,3}} = \hat{\vec{k}}_3$ and
\begin{equation}
  \hat{\vec{x}}_{\mathrm{loc,3}} =
   \begin{bmatrix}
     \cos\psi \\ \sqrt{2/3}\sin\psi \\ \sqrt{1/3}\sin\psi \\ 
   \end{bmatrix}, \qquad 
  \hat{\vec{y}}_{\mathrm{loc,3}} =
   \begin{bmatrix}
     -\sin\psi \\ \sqrt{2/3}\cos\psi \\ \sqrt{1/3}\cos\psi
   \end{bmatrix}.
\end{equation}
Next, provided that the resonator vacuum field $\delta\vec{B}$ is
known, the coupling constants $g_x$ and $g_y$ for the two NV center
transitions follow as $g_x = \gNV\muB\delta\vec{B}\cdot
\hat{\vec{x}}_{\mathrm{loc}}$ and $g_y = \gNV\muB\delta\vec{B}\cdot
\hat{\vec{y}}_{\mathrm{loc}}$ for each spin family. The sign (or in
general the phase) of $g$ is irrelevant and we use in the simulations
a positive quantity for the coupling constants. For the spin families
1 and 3 we then find:
\begin{equation}
  \begin{split}
      |g_{x,1}| &= \frac{\gNV\muB|\cos\psi|}{\sqrt{3}}
     |\delta B_x  - \sqrt{2}\delta B_y|, \\
  |g_{y,1}| &= \frac{\gNV\muB|\sin\psi|}{\sqrt{3}}
     |\delta B_x - \sqrt{2}\delta B_y|, \\
  |g_{x,3}| &= \gNV\muB \left|\cos\psi \cdot\delta B_x 
     + \sqrt{\frac{2}{3}}\sin\psi\cdot\delta B_y\right|, \\
  |g_{y,3}| &= \gNV\muB \left|\sin\psi\cdot\delta B_x 
    - \sqrt{\frac{2}{3}}\cos\psi\cdot\delta B_y\right|.
  \end{split}
\end{equation}
The distribution of $g$-parameters thus follows from both an
inhomogeneous distribution of magnetic fields $\delta\vec{B}$ and the
angular distribution of crystal strain experienced by the NV
centers. We note that the distributions of $|g_{x,1}|$ and $|g_{y,1}|$
(and likewise for $|g_{x,3}|$ and $|g_{y,3}|$) are identical when the
angular average is taken into account.

\begin{figure}[t]
  \centering
  \includegraphics[width = \linewidth]{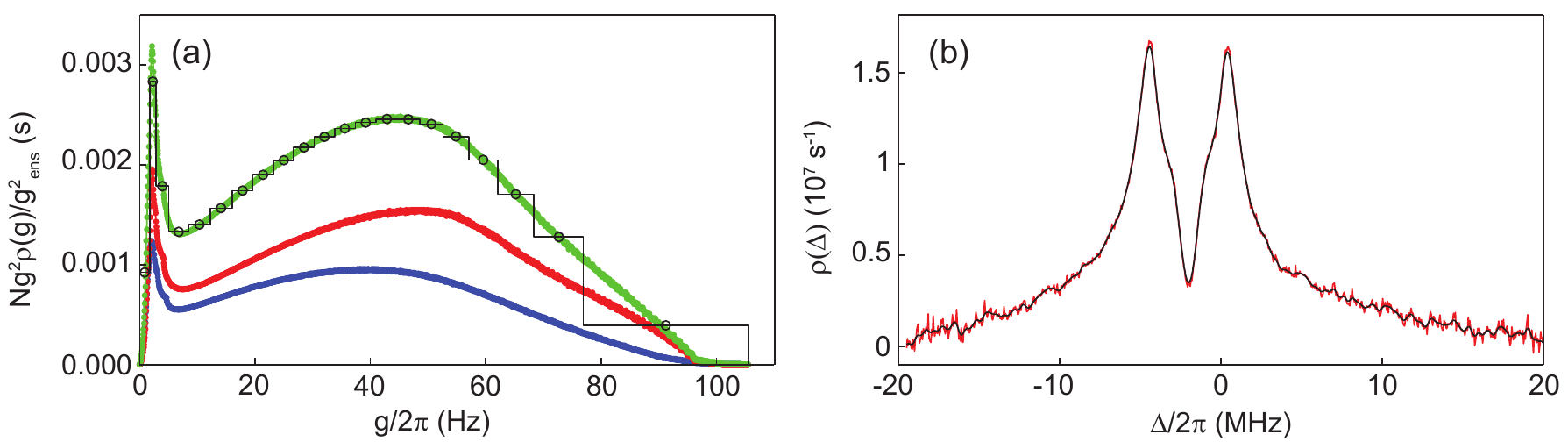}  
  \caption{(a) The distribution of coupling strengths $\rho(g)$
    plotted as $g^2\rho(g)$ and normalized to unity area
    [$\int_0^{\infty}\rho(g)dg = 1$]. From below: The non-orthogonal
    families 1 + 2 (blue), the orthogonal families 3 + 4 (red), and
    the total contribution from all families (green). The black
    circles denote the discrete distribution from the multi-mode
    simulations with $M_g = 21$ bins. (b) The coupling-density profile
    $\rho(\Delta)$ versus spin frequency
    [$\int_{-\infty}^{\infty} \rho(\Delta)d\Delta =
    \gens^2$]. The red curve is extracted experimentally while the
    black curve is a smoothed version with $M_{\Delta} = 3001$ bins
    used for the multi-mode simulations.}
  \label{fig:ExtractDist}
\end{figure}

To proceed, the vacuum field $\delta\vec{B}$ is calculated using the
COMSOL simulation software. The shape of $\delta\vec{B}$ is shown graphically in the
inset of Fig.~1b in the main text, and the curved sections of the
resonator element was neglected such that $\delta\vec{B}$ essentially
only consists of an $x$- and a $y$-component. The correct magnitude
for $\delta\vec{B}$ is obtained by scaling the field to the one
corresponding to the resonator current set equal to $\delta I =
\omegac\sqrt{\hbar/2Z_0}$. The resulting distribution [or rather $g^2$ times the
distribution $\rho(g)$ with $\int_0^{\infty} \rho(g)dg = 1$] of
$g$-parameters is shown in Fig.~\ref{fig:ExtractDist}a, where the
effective length of the active crystal was taken to be 100 microns
along the $z$-axis, the concentration was 2 ppm, and a 0.7 $\mu \mathrm{m}$
spacing for glue was estimated between the resonator and the
diamond crystal. The area under $Ng^2\rho(g)$ is equal to the squared
ensemble coupling constant $\gens^2$, and the orthogonal families (3
and 4) contribute $\frac{5}{8}$ of the total area, while families 1
and 2 contribute the remaining $\frac{3}{8}$ part. The black circles
in Fig.~\ref{fig:ExtractDist}a denote the actual discrete
distribution, which was used for the simulation results shown in Fig. 4b,c,e and 5b in the main text.

\paragraph{Determining the frequency distribution}

The frequency distribution $\rho(\omega)$ is determined as explained earlier, from microwave absorption measurements. Note that the directly measured ensemble coupling constant from integrating
the curves in Fig.~\ref{fig:ExtractDist}b is $2\pi\cdot 5.0$
MHz. However, these data were obtained under repumping with a laser power of $180 \mu \mathrm{W}$ corresponding (see Fig.~3b of main text) to a spin polarization $p =
-S_z/N = 0.64$. In the simulations we thus rescale the number of spins
$N$ such that the ensemble coupling constant becomes $\gens =
2\pi\cdot 5.0\:\mathrm{MHz}/\sqrt{p} = 2\pi\cdot 6.3\:\mathrm{MHz}$,
and when simulating the spin-echo sequences the appropriate initial
value of $S_z = -Np'$ is chosen, where $p'$ refers to the polarization
corresponding to the laser power used in the experiment ($1.5$\,mW in most measurements).

We note that the $g$-parameter distribution calculated from first
principles as explained above leads to $\gens =
2\pi\cdot 4.4$ MHz, i.e.~$\approx$ 70 \% of the value stated
above. Given the measurement accuracy, and the fact that the NV center
concentration and the resonator-to-crystal distance are known only approximately, we find the
agreement between these numbers satisfactory.

\paragraph{Retrieval efficiency for the multi-mode echo sequence}
\label{sec:fidel-multi-mode}
The main purpose of our simulations is to understand the measured retrieval efficiency of the echo pulses. Hence, we must be quantitative on the
echo-pulse magnitudes in the simulations, and accordingly we must
adapt the equations of motion to cover the actual case with two spin
classes of different coherence time as explained in Sec. IV.

\begin{figure}[t]
  \centering
  \includegraphics[width=\linewidth]{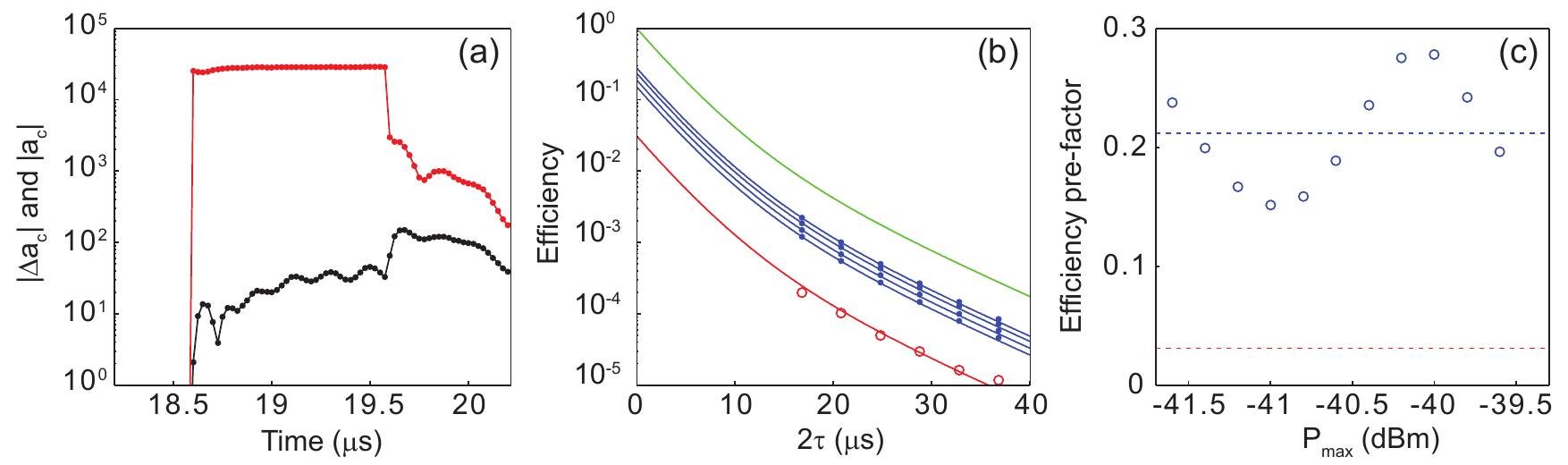}
  \caption{(a) The resonator field mean value $|a_{\mathrm{c}}|$
    during the refocusing pulse (red) compared to the difference,
    $|\Delta{a}_{\mathrm{c}}| = |a_{\mathrm{c},a} - a_{\mathrm{c},b}|$
    (black), between the simulations with fast and slow
    dephasing. The refocusing pulse extends from $18.6$--$19.6$ \micro{s},
    followed by free-induction decay. (b) The red circles denote the
    field recovery efficiency of the experimental echo pulses versus the delay $2\tau$
    between storage and readout of the six pulses in the multi-mode
    echo sequence. The blue dots are simulated efficiencies for a range
    of driving powers in the vicinity of the experimental value of
    $-40.6$ dBm. The green solid line is the square of the function
    $f(\tau) = A \exp(-2\tau/T_{2A}) + B\exp(-2\tau/T_{2B})$
    described in the main text. The remaining solid lines corresponds
    to a constant pre-factor times $[f(\tau)]^2$. (c) The blue circles
    correspond to the simulated pre-factors from panel (b) with the
    horizontal blue dashed line denoting the average value of
    $0.21$. The red dashed line is the experimental pre-factor equal
    to $0.031$.}
  \label{fig:FidelityCalculation}
\end{figure}

The obvious strategy would be doubling the sub-ensemble partitioning
into two parts---one with coherence time $T_{2A}$ and the other with
$T_{2B}$---and adapting
Eqs.~(\ref{eq:def_sub-ensembles})-(\ref{eq:Mean_val_eq_Sz_kl})
accordingly. However, we use a much simpler strategy in the following:
The simulations are run first with a single coherence time $T_{2A}$,
and then repeated with the other coherence time $T_{2B}$. The
calculated reflected fields in these two instances are then combined
according to the weights given by the fitting parameters $A$ and
$B$ mentioned in the main text. Now the two sub-ensembles share in practice a common resonator field
$\ac$ which in the simulations is ascribed two different values, and
under influence of the refocusing pulse, i.e.~in conditions of saturation,
linearity will not be applicable. However, for our problem in question
(the multi-mode echo sequence in Fig.~5 of the main text) the refocusing pulse
strength is $\approx 15$ dB above saturation, and hence the external
driving determines the resonator field to a much larger extent than
the reaction field of the spin dipoles. Indeed,
Fig.~\ref{fig:FidelityCalculation}a shows that the resonator field is
essentially identical in the two simulation runs with fast and slow
coherence time.

Now, for the multi-mode sequence shown in Fig.~5 in the main text, the
efficiency is calculated for each of the six pulses in the sequence and
plotted as a function of the input-output delay $2\tau$ in
Fig.~\ref{fig:FidelityCalculation} (red circles). Since the efficiency
is defined as the energy of the echo pulse relative to the incoming
energy,it is expected to behave as : $\mathrm{Efficiency} =
c\cdot [f(\tau)]^2$, where $c$ is a constant and $f(\tau)$ was found
in the main text as fitting the experimental amplitude-decay of echo
pulses [scaled such that $f(\tau = 0) = 1$]. The function
$[f(\tau)]^2$ is shown as the green curve in
Fig.~\ref{fig:FidelityCalculation}b, and this curve corresponds to the
efficiency that one would obtain if spin dephasing were the sole reason
for a non-ideal echo protocol. We observe that both simulations (blue
dots) and experiment lead to a lower efficiency; however, the overall
behavior corresponds indeed to a pre-factor $c$ multiplied onto
$[f(\tau)]^2$.

Before we comment on these pre-factors, there is a little subtlety to
mention about the simulations: Due to the discrete division of the
coupling strengths into $M_g = 21$ bins (see
Fig.~\ref{fig:ExtractDist}a), there are artificial oscillations
occurring in the echo recovery efficiency as a function of the applied driving
power, see the blue circles in Fig.~\ref{fig:FidelityCalculation}c. We
have checked that an increasing $M_g$ will decrease the magnitude of
such oscillations (not shown) while the mean value stays essentially
fixed. For this reason we compute the mean value of the efficiency
pre-factors in Fig.~\ref{fig:FidelityCalculation}c, which then amounts
to 0.21 (dotted blue line). We thus conclude that the rather low
simulated efficiencies (a few times $10^{-3}$ and decreasing with
increasing $2\tau$ in Fig.~\ref{fig:FidelityCalculation}b) are caused
primarily by spin dephasing. The simulated curve
in Fig.~4e of the main text was obtained with $M_g = 50$.

Next, the experimental retrieval efficiencies follow also the trend of
$[f(\tau)]^2$ but with a pre-factor $\approx 7$ times lower than the
one found from simulations. This indicates that there is an
additional effect in play, which is not included in our
simulations. As explained in the main text, we attribute this discrepancy to the inadequacy of the Markov approximation to describe decoherence caused by a spin bath.

% Uncomment below to make bibliography. 
%

\end{document}